\documentclass[11pt,a4paper]{article}
\usepackage{jheppub}
\usepackage{natbib}

\usepackage[normalem]{ulem}

\usepackage{amsmath,amssymb,amsthm,bm,braket,ascmac,bbm}
\usepackage{graphicx}
\usepackage{hyperref}
\usepackage{slashed}

\title{
Can we explain cosmic birefringence without a new light field 
beyond Standard Model?
}

\author[a,b]{Yuichiro Nakai}
\author[c]{Ryo Namba}
\author[d,e]{Ippei Obata}
\author[a,b]{Yu-Cheng Qiu}
\author[e,f]{Ryo Saito}

\affiliation[a]{
Tsung-Dao Lee Institute, Shanghai Jiao Tong University, \\ 520 Shengrong Road, Shanghai 201210, China
}
\affiliation[b]{
School of Physics and Astronomy, Shanghai Jiao Tong University, \\ 800 Dongchuan Road, Shanghai 200240, China
}
\affiliation[c]{
RIKEN Interdisciplinary Theoretical and Mathematical Sciences (iTHEMS),
}
\affiliation[d]{
Max-Planck-Institut f{\"u}r Astrophysik, Karl-Schwarzschild-Str. 1, 85748 Garching, Germany,
}
\affiliation[e]{
Kavli Institute for the Physics and Mathematics of the Universe, \\
Todai Institute for Advanced Study, The University of Tokyo, \\
Chiba 277-8583, Japan (Kavli IPMU, WPI)
}
\affiliation[f]{
Graduate School of Science and Engineering, \\
Yamaguchi University, Yamaguchi 753-8512, Japan
}

\abstract{
The recent analysis of the Planck 2018 polarization data shows a nonzero isotropic cosmic birefringence (ICB)
that is not explained within the $\Lambda$CDM paradigm.
We then explore the question of whether the nonzero ICB is interpreted
by the framework of the Standard Model Effective Field Theory (SMEFT),
or at the energy scales of the cosmic microwave background,
the low-energy EFT (LEFT) whose dynamical degrees of freedom are five SM quarks
and all neutral and charged leptons.
Our systematic study reveals that any operator in the EFT on a cosmological background 
would not give the reported ICB angle,
which is observationally consistent with frequency independence.
In particular, we estimate the size of the ICB angle generated
by the effect that the cosmic microwave background photons travel
through the medium of the cosmic neutrino background with
parity-violating neutrino-photon interactions and find that it would be too small to explain the data.
If the reported ICB angle should be confirmed,
then our result would indicate the existence of a new particle that is lighter than the electroweak scale and
feebly interacting with the SM particles. 
}

\begin{document}

\maketitle

\section{Introduction}

Precision measurements of the cosmic microwave background (CMB) radiation play a central role in modern cosmology
and enable us to deepen our understanding of the fundamental laws of nature.
The Universe looked through the eyes of the WMAP and {\it Planck} satellites
is well-fitted by the $\Lambda$ Cold Dark Matter ($\Lambda$CDM) model~\cite{WMAP:2003elm,Komatsu:2014ioa,Planck:2013pxb,Planck:2018vyg}
which has long become a cornerstone of the standard cosmology.
However, the recent analysis of CMB polarization data has
measured a parity-violating signal~\cite{Minami:2020odp,Diego-Palazuelos:2022dsq,Eskilt:2022wav,Eskilt:2022cff,Eskilt:2023ndm}, called {\it cosmic birefringence}~\cite{Carroll:1989vb,Carroll:1991zs,Harari:1992ea}, 
which may show us a hint of new physics beyond the $\Lambda$CDM paradigm.

Cosmic birefringence is a phenomenon that rotates the plane of linear polarization of the CMB photons. 
Its overall rotation angle from the last scattering surface to the present, called isotropic cosmic birefringence (ICB) angle and hereafter denoted by $\beta$, has been probed in the past CMB measurements \cite{Feng:2006dp,QUaD:2008ado,WMAP:2010qai,Planck:2016soo}.
However, the uncertainty of systematic error from the instrumental miscalibration of the polarization angle has strongly limited the determination of $\beta$.
To overcome this issue, the method relying on the polarized Galactic foreground emission to extract the intrinsic effect on $\beta$ has been developed~\cite{Minami:2019ruj,Minami:2020xfg,Minami:2020fin}, and ref.~\cite{Minami:2020odp} has recently reported a nonzero ICB angle of $\beta = 0.35^\circ \pm 0.14^\circ $ ($2.4\sigma$) at the $68\%$ confidence level for nearly full-sky {\it Planck} polarization data.
The precision of $\beta$ has been improved and the latest joint analysis of {\it Planck}/WMAP data has reported $\beta = 0.34^\circ  \pm 0.09^\circ$
($3.6\sigma$)~\cite{Eskilt:2022cff}.
Moreover, the measured $\beta$ is consistent with frequency independence and does not favor a possibility of Faraday rotation effect caused by the local magnetic field \cite{Eskilt:2022wav}.
Although the contribution to a systematic error in the ICB angle from Galactic foregrounds is not yet well understood~\cite{Clark:2021kze,Diego-Palazuelos:2022cnh}, we could avoid this problem by developing the method that does not rely on the foreground contribution but reduces the impact of the miscalibration angle in the upcoming CMB observations~\cite{Monelli:2022pru,Jost:2022oab} (see ref.~\cite{Komatsu:2022nvu} for a detailed review).
Therefore, we can expect a solid confirmation of the nonzero ICB angle in the near future,
and it is timely to explore its origin.

The measured ICB can be caused when the CMB photons pass through a cosmological background of a pseudoscalar field $\phi$ that is weakly coupled to the photon
through a Chern-Simons (CS) term $\phi F_{\mu\nu} \tilde{F}^{\mu\nu}$,
where $F$ and $\tilde{F}$ respectively denote the electromagnetic tensor and its dual \cite{Carroll:1998zi,Lue:1998mq}. 
A prevailing candidate of the pseudoscalar field $\phi$ has been provided by an axion-like particle (ALP),
and a possibility that photon's birefringence is caused by a cosmic background of the ALP
constituting dark energy or dark matter has been developed~\cite{Pospelov:2008gg,Finelli:2008jv,Panda:2010uq,Lee:2013mqa,Zhao:2014yna,Liu:2016dcg,Sigl:2018fba,Fedderke:2019ajk}.
Then, after the measurement of the nonzero ICB has been reported, most of the previous studies have focused on the implications for the ALP~\cite{Fujita:2020ecn,Takahashi:2020tqv,Fung:2021wbz,Nakagawa:2021nme,Jain:2021shf,Choi:2021aze,Obata:2021nql,Nakatsuka:2022epj,Lin:2022niw,Gasparotto:2022uqo,Lee:2022udm,Jain:2022jrp,Murai:2022zur,Gonzalez:2022mcx,Qiu:2023los,Eskilt:2023nxm,Namikawa:2023zux,Gasparotto:2023psh}.
However, to explain the reported ICB, the ALP mass should be extremely light~\cite{Fujita:2020ecn}. 
The existence of such an ultra-light ALP has significant implications for physics beyond the Standard Model (SM), {\it e.g.,} ruling out simple Grand Unified models~\cite{Agrawal:2022lsp}. 
Then, one would wonder whether there are other possibilities to generate the ICB or not.
In the present paper, 
as a step to identify new physics behind the reported nonzero ICB, 
we explicitly show that it requires at least a new light particle other than the SM particles under the standard cosmological evolution. 

When a new light particle is absent, an operator that induces the ICB should be written only in terms of the SM fields. 
Then, the SM effective field theory (SMEFT) 
and low-energy effective field theory (LEFT) provide 
a powerful tool to systematically list up
all such operators 
in the SM and its extensions. 
The SMEFT includes all operators of the SM fields that respect the gauge symmetry $SU(3)_{C} \times SU(2)_{L} \times U(1)_{Y}$.
This framework includes any possible 
explanations of the ICB with the SM fields known in the literature 
including scatterings between photons and fermions~\cite{Bartolo:2019eac}. 
As we will see, it is also convenient to introduce the so-called LEFT, 
the EFT below the electroweak breaking scale. 
In the LEFT, 
we assume that
there are no new particles
(other than possible light sterile neutrinos)
around or below the electroweak scale 
and the interactions respect the gauge symmetry $SU(3)_{C} \times U(1)_{\rm EM}$. 

We report the following two results: 
(i) only a CS-type effective operator, $\tilde{\cal O} F_{\mu\nu} \tilde{F}^{\mu\nu}$ with a Lorentz-scalar operator $\tilde{\cal O}$,
is able to induce the frequency-independent ICB in our Universe,
and 
(ii) none of the CS-type effective operators in the SMEFT/LEFT leads to the desired ICB angle. 
We note that the operator $\tilde{\cal O} F_{\mu\nu} \tilde{F}^{\mu\nu}$ is distinguished from $J_\mu A_\nu \tilde{F}^{\mu\nu}$ with a vector current $J_\mu$. 
It has been reported that
the effective operator $J_\mu A_\nu \tilde{F}^{\mu\nu}$ for the neutrino current appears via the loop interactions between photon and neutrino, and leading to the photon's birefringence \cite{Royer:1968rg,Karl:1975df,Mohanty:1997mr,Karl:2004bt,Dvornikov:2020olb,Petropavlova:2022spq}.
However, the operator $J_\mu A_\nu \tilde{F}^{\mu\nu}$ for the neutrino current is not solely gauge invariant under $U(1)_{\rm EM}$ 
and hence does not appear within SMEFT/LEFT.\footnote{This claim would not be true when a photon has an effective mass in a plasma or two lepton loop diagrams are considered. 
However, this effect usually gives rise to a very tiny birefringence angle \cite{Dvornikov:2020olb} or frequency-dependent birefringence angle \cite{Karl:2004bt}.}
In the study of cosmic birefringence by this operator, a couple of scenarios beyond SMEFT/LEFT have been developed to get a sizable amount of ICB angle \cite{Geng:2007va, Ho:2010aq, Zhou:2023aqz}.
Therefore,
if the data should be confirmed,
our results would then indicate the breakdown of the SMEFT/LEFT and thus 
the existence of a new particle lighter than the electroweak scale. 

The rest of the paper is organized as follows.
Section~\ref{s:relevant op} shows that only CS-type operators are relevant to the generation of the frequency-independent ICB.
In section~\ref{CS_SMEFT}, we list up all the possible CS-type operators in the SMEFT/LEFT.
Then, section~\ref{ICB} discusses whether the listed CS-type operators can induce the reported nonzero ICB with the corresponding cosmological backgrounds.
In section~\ref{s:extensions}, we extend our arguments to narrow down possible new particles that can explain the reported nonzero ICB.
Section~\ref{conclusion} is devoted to conclusions.
Some calculational details are summarized in appendices.

\section{Operators relevant to ICB}
\label{s:relevant op}

Let us first show that interactions relevant to the frequency-independent ICB are only given by CS-type operators. 
The ICB requires an effective parity-violating operator quadratic in the photon field $A_\mu$ because it is caused by a difference between the phase velocities of the left- and right-handed photons. 
In the vacuum, due to the Lorentz and $U(1)_{\rm EM}$ symmetries, 
the effective quadratic action of the photon field $A_\mu$ is only described by the operator,
    \begin{align}
        F_{\mu\nu} F^{\mu\nu} \;,
    \end{align}
with the field-strength tensor $F_{\mu\nu} \equiv \nabla_\mu A_\nu - \nabla_\nu A_\mu$. 
The parity-violating CS operator,
    \begin{align}
        F_{\mu\nu} \tilde{F}^{\mu\nu} \,; \quad \tilde{F}^{\mu\nu} \equiv \epsilon^{\mu\nu\alpha\beta} F_{\alpha\beta} / 2 \;,
    \end{align}
is a total-derivative term, when entering in the action with a constant coefficient, and thus does not contribute to the local action. 
Note that $\epsilon^{\mu\nu\rho\sigma} = \eta^{\mu\nu\rho\sigma} / \sqrt{-g}$, where we choose the convention that the flat-space Levi-Civita symbol takes $\eta^{0123} = 1$.
Since there is no parity-violating term in the action, 
the ICB is not generated in the vacuum: it requires a medium. 

The isotropy of the measured rotation angle indicates that the medium is homogeneous over the Universe\footnote{If the medium is inhomogeneous, the birefringence angle has random contributions from different coherent patches of the medium along each line of sight. It would be very rare that the birefringence angles for different directions coincide with each other.}
and thus made up of stable matter with a cosmological background. 
Moreover, the charged components can be excluded from our considerations.
The stable charged SM particles are only electrons and protons. 
Their contributions are negligible compared to the neutral ones because the cosmological backgrounds of electrons and protons are suppressed by the small baryon-to-photon ratio.
Therefore, we can assume that the medium is homogeneous and neutral.
Under the standard cosmological evolution, 
the candidate matter that constitutes the medium is limited to the following: the Higgs vacuum expectation value (VEV), the quark pair/gluon condensates, the cosmic neutrino background (C$\nu$B), and the cosmological magnetic field.%
\footnote{We shall omit gravitational effects on ICB in the following considerations. Beyond minimal couplings between gauge fields and gravity, it has been shown that there exists a unique non-minimal coupling without pathology of the form $\tilde{R}_{\mu\nu\rho\sigma} F^{\mu\nu} \tilde{F}^{\rho\sigma}$, where $\tilde{R}_{\mu\nu\rho\sigma}$ is the dual of the Riemann tensor \cite{Horndeski:1976gi}. Not only is the curvature tensor of the order of $H^2$ ($H$: Hubble parameter) on the cosmological background, but this term does not break parity, thus no contribution to ICB. }

The presence of a cosmological background can break some symmetries in the effective quadratic action of the photon field and allow operators other than $F_{\mu\nu}F^{\mu\nu}$ 
when the fields are expanded around the background. 
From the neutrality of the background, 
the effective action should respect the $U(1)_{\rm EM}$ symmetry. 
Let us also assume that the background respects the cosmological principle: 
The universe looks homogeneous and isotropic in the CMB rest frame. 
This is valid except for the cosmological magnetic field among the candidates. 
Then, in the CMB rest frame, the effective action should have spatial invariance. 
From the gauge and spatial invariance, 
the effective quadratic action should have the following operators: 
    \begin{align}\label{op:EFT spatial}
        c_{EE} |{\bf E}|^2 + c_{BB} |{\bf B}|^2 + c_{EB} {\bf E} \cdot {\bf B} \;,
    \end{align}
and operators with derivative(s) on ${\bf E}$ and ${\bf B}$, 
where the coefficients are functions of the cosmic time $t_c$. 
Here, the electric field ${\bf E}$ and the magnetic field ${\bf B}$ are defined in the CMB rest frame. 
The functions $c_{EE}$ and $c_{BB}$ correspond to an electric permittivity and a magnetic permeability, which deviate from those in the vacuum due to the presence of the cosmological medium. 
Now, we also have the parity-violating term ${\bf E} \cdot {\bf B}$ in the action. 
Since the coefficient $c_{EB}$ is a function of the cosmic time $t_c$, 
it is not reduced to the total derivative term. 
In appendices~\ref{a:macro} and~\ref{a:propagation medium}, 
we show that the operator ${\bf E} \cdot {\bf B}$ actually induces the ICB. 
The operators with the derivative(s) give a frequency-dependent ICB angle, 
which is inconsistent with observations \cite{Eskilt:2022wav}. 
Therefore, we do not consider them further. 
This argument indicates that any operator relevant to the reported ICB should be reduced to the ${\bf E} \cdot {\bf B}$ term in a cosmological background. 

It has not been proved yet that only the CS-type operator leads to the frequency-independent ICB. 
There might be operators other than $F_{\mu\nu} \tilde{F}^{\mu\nu}$ that reduce to ${\bf E} \cdot {\bf B}$ in a cosmological background. 
First, we consider parity-violating operators of the form, 
    \begin{align}\label{op:pv ff}
        J_{\alpha\beta\mu\nu}F^{\alpha\beta}\tilde{F}^{\mu\nu} \,,
    \end{align}
where $J_{\alpha\beta\mu\nu}$ is a tensor with even parity written in terms of the metric and matter fields. 
From the cosmological principle, the cosmological background of $J_{\alpha\beta\mu\nu}$ should be written in terms of the metric $g_{\mu\nu}$ and the unit four-vector $u_\mu \propto \nabla_\mu t_c$ for the cosmic time $t_c$ because only these two tensors are parity-even and invariant under the spatial rotation in the CMB rest frame where $u_\mu \propto \delta_\mu^0$. 
If $J_{\alpha\beta\mu\nu}$ does not contain $g_{\mu\nu}$, 
$J_{\alpha\beta\mu\nu}$ should be proportional to $u_{\alpha}u_{\beta}u_{\mu}u_{\nu}$ and the operator (\ref{op:pv ff}) vanishes. 
Therefore, 
$J_{\alpha\beta\mu\nu}$ should have the following form, 
    \begin{align}
        J_{\alpha\beta\mu\nu} =  \frac{1}{4}(g_{\alpha\mu}J_{\beta\nu} + g_{\beta\nu}J_{\alpha\mu} - g_{\beta\mu}J_{\alpha\nu} - g_{\alpha\nu}J_{\beta\mu}) \,,
    \end{align}
and the operator (\ref{op:pv ff}) reduces to
    \begin{align}\label{op:pv example}
        J_{\alpha\beta} F^{\alpha \mu} \tilde{F}^{\beta}{}_{\mu} \,,
    \end{align}
in the cosmological background. 
We can show that this operator is rewritten in the CS form 
using the identity \cite{Fleury:2014qfa},
\begin{align}
\label{eq:J identity}
    F_{\alpha\mu} \tilde{F}^{\beta\mu} = \frac{\delta_\alpha^\beta}{4} \, F_{\mu\nu} \tilde{F}^{\mu\nu} \; .
\end{align}
In terms of $J_{\alpha\beta\mu\nu}$, the resultant CS-type operator is given by
\begin{align}
\label{eq:J CS}
     \tilde{\cal O} F_{\mu\nu} \tilde{F}^{\mu\nu} \;; \quad \tilde{\cal O} = \frac{J_\mu{}^\mu}{4} = \frac{J_{\alpha\beta}{}^{\alpha\beta}}{6} \,.  
\end{align}
Therefore, it would be enough to study the CS-type operators for the Lorentz scalar $\tilde{\cal O} = J_{\alpha\beta}{}^{\alpha\beta}/6$. 

Another possibility is
    \begin{equation}\label{op:jk}
        J_\mu K^\mu \;; ~ K^\mu \equiv 2 A_\nu \tilde{F}^{\mu\nu} \;,
    \end{equation}
for the CS current $K^\mu$ ($\nabla_\mu K^\mu = F_{\mu\nu}\tilde{F}^{\mu\nu}$) and a matter field $J_\mu$. 
This operator is reduced to $c_{\rm EB} {\bf E} \cdot {\bf B}$ with $\dot{c}_{\rm EB} = J_0/4$. 
However, it does not respect $U(1)_{\rm EM}$ in general. 
To achieve the invariance under the gauge transformation $A_{\mu} \to A_{\mu} + \nabla_\mu \alpha$, 
the current $J_\mu$ should identically satisfy the integrability condition $\nabla_{[\mu} J_{\nu]} = 0$,
{\it i.e.,} $J_\mu = \nabla_\mu \tilde{\cal O}$ for a Lorentz scalar $\tilde{\cal O}$. 
Then, $J_\mu K^\mu$ can be rewritten in the CS form $\tilde{\cal O} F_{\mu\nu}\tilde{F}^{\mu\nu}$ with a partial integration. 
Note that $J_\mu K^\mu$ can be embedded into the gauge invariant operator
    \begin{equation}\label{eq:ffF}
        \bar{\psi}\gamma_\mu D_\nu \psi \tilde{F}^{\mu\nu} \;,
    \end{equation}
with $J_\mu = \bar{\psi}\gamma_\mu \psi$ for a fermion field $\psi$ and $D_\mu$ its covariant derivative. However, the fermion field should be electrically charged and this option is excluded from the neutrality of the background. 
In general, to get the interaction (\ref{op:jk}), 
our model should contain a field playing the role of a Stueckelberg field or Nambu-Goldstone (NG) field such as the phase of an electrically-charged field in the interaction (\ref{eq:ffF}).
It has been shown in \cite{Geng:2007va, Ho:2010aq, Zhou:2023aqz} 
that the interaction (\ref{op:jk}) can induce a sizable amount of cosmic birefringence with employing a two-form field as a Stueckelberg field.
In our setup, a candidate for a Stueckelberg field or NG field is absent for $A_\mu$. 
Therefore, we do not need to consider the interaction (\ref{op:jk}) (see also the discussion in section \ref{s:extensions}). 
%


Finally, we consider an operator in the form
    \begin{align}\label{op:jf}
        J_{\mu\nu} F^{\mu\nu} \,.
    \end{align}
This operator vanishes when we replace the matter field $J^{\mu\nu}$ by the cosmological background, which should be written in terms of $g_{\mu\nu}$ and $u_\mu$. 
However, it can affect the propagation of CMB photons when their backreaction to the cosmological background is taken into account. 
Schematically, we can write the backreaction term as
    \begin{align}\label{eq:response}
        \delta J_{\mu\nu} = \hat{K}_{\mu\nu\alpha\beta} F^{\alpha\beta} \,,
    \end{align}
by separating $J_{\mu\nu}$ into the background and the deviation induced by a propagating photon, $\delta J_{\mu\nu}$. 
Here, the response function $\hat{K}_{\mu\nu\alpha\beta}$ is a non-local operator in general. 
Then, substituting Eq.~(\ref{eq:response}) to $J_{\mu\nu} F^{\mu\nu}$, 
we obtain the effective operator
    \begin{align}\label{op:KCS}
        F^{\mu\nu} \hat{K}_{\mu\nu\alpha\beta} F^{\alpha\beta} \,.
    \end{align}
If the response function $\hat{K}_{\mu\nu\alpha\beta}$ has a component
    \begin{align}
        \hat{K}_{\mu\nu\alpha\beta} \supset \hat{\cal O}_{\epsilon} \epsilon _{\mu\nu\alpha\beta}/2 \,,
    \end{align}
the operator (\ref{op:KCS}) results in an operator
    \begin{align}\label{op:OeCS}
        F_{\mu\nu} \hat{\cal O}_{\epsilon} \tilde{F}^{\mu\nu} \,.
    \end{align}
In general, $\hat{\cal O}_{\epsilon}$ is a non-local operator and thus the operator (\ref{op:OeCS}) induces a frequency-dependent ICB angle. 
In appendix \ref{dipole op}, 
we will show that the resultant ICB angle actually depends on the frequency in the framework of SMEFT/LEFT. 
Therefore, we do not need to consider the interaction (\ref{op:jf}).\footnote{Ref.~\cite{Khodagholizadeh:2023aft} has developed a model of dipole interaction between CMB photon and dark matter and has also shown that it is hard to explain the observed ICB angle.}


The above discussion is not applied in the presence of a cosmological magnetic field $\bar{\bf B}$
because it breaks the isotropy to the axial symmetry along the direction. 
As we have done in the isotropic case of Eq.~(\ref{op:EFT spatial}), 
it is convenient to work in the CMB rest frame with $u^\mu \propto \delta^\mu_0$ 
to write down the most general operators. 
Here, the homogeneity requires that $|\bar{\bf B}|$ should be a function of the cosmic time $t_c$. 
Then, the kinetic action is given by quadratic terms in the propagating photon fields, ${\bf E}_\parallel$, ${\bf E}_\perp$, ${\bf B}_\parallel$ and ${\bf B}_\perp$, where the subscripts $\parallel$ and $\perp$ respectively denote the components parallel and orthogonal to 
the background magnetic field. 
Since the background magnetic field is an axial vector, 
the parallel and orthogonal components respectively
have the opposite and same parity as the original vector. 
Thus, the kinetic action consists of the parity-violating operators
    \begin{align}\label{op:aniso pv}
        {\bf E}_\parallel \cdot {\bf B}_\parallel \,, \quad {\bf E}_\perp \cdot {\bf B}_\perp \,,
    \end{align}
as well as the parity-conserving operators
    \begin{align}\label{op:aniso pc}
        {\bf E}_\parallel \cdot {\bf E}_\parallel \,, \quad {\bf E}_\perp \cdot {\bf E}_\perp\,, \quad
        {\bf B}_\parallel \cdot {\bf B}_\parallel \,, \quad {\bf B}_\perp \cdot {\bf B}_\perp \,.
    \end{align}
The latter parity-conserving terms can be interpreted as that permittivity and permeability tensors become non-diagonal due to the anisotropic medium. 
As for the parity-violating terms, 
in addition to the CS form $F_{\mu\nu}\tilde{F}^{\mu\nu} \propto {\bf E} \cdot {\bf B}$, 
new independent parity-violating operators ${\bf E}_\parallel \cdot {\bf B}_\parallel$ can appear (an ${\bf E}_\perp \cdot {\bf B}_\perp$ term can be rewritten in terms of $F_{\mu\nu} \tilde{F}^{\mu\nu}$ and ${\bf E}_\parallel \cdot {\bf B}_\parallel$). 
As we will give proof in appendix~\ref{a:propagation medium}, this new operator generates the anisotropic cosmic birefringence (ACB). 
\footnote{ACB can be a cosmological probe of new physics and a target of future CMB experiments. See, {\it e.g.,} refs. \cite{Greco:2022ufo, Greco:2022xwj}.}
It should be also noted that the parity-conserving operators (\ref{op:aniso pc}) can cause a cosmic birefringence in the anisotropic background 
because the cosmological magnetic field $\bar{\bf B}$ can modify the dispersion relations for the polarization modes parallel and orthogonal to $\bar{\bf B}$. 
However, as expected from the fact that the modification depends on the relative angle of the propagating direction with $\bar{\bf B}$, the resultant cosmic birefringence is anisotropic (see ref.~\cite{Brezin:1971nd}). 
Moreover, it depends on the frequency, which is not consistent with the report in ref.~\cite{Eskilt:2022wav}. 
Therefore, we conclude that only the CS-type operators $\tilde{\cal O} F_{\mu\nu} \tilde{F}^{\mu\nu}$ can generate the reported frequency-independent ICB. 

\section{CS-type operators in SMEFT/LEFT}
\label{CS_SMEFT}

We now list up CS-type operators in the SMEFT or LEFT,
\begin{equation}
    \mathcal{L}_{\rm CS} = \frac{\alpha}{8\pi} \sum_a \frac{\tilde{\mathcal{O}}_a}{\Lambda_a^n} \, F_{\mu\nu} \tilde{F}^{\mu\nu} \; ,
    \label{eq:L_CS}
\end{equation}
where the subscript $a$ denotes the operator species, $\Lambda_a$ is some mass scale
and the power $n$ is given by the dimension of the operator $\tilde{\mathcal{O}}_a$. 
We have factored out the electromagnetic constant $\alpha / (8\pi)$ as a convention. 
As shown in the previous section, any particle processes that might lead to the ICB
should be described by Eq.~\eqref{eq:L_CS} as an effective Lagrangian. 
We henceforth list possible $\tilde{\mathcal{O}}_a$ of each dimension in the SMEFT/LEFT. 
\\


{\bf Dimension 2 --} Let us first write down effective operators $\tilde{\mathcal{O}}_a$ of dimension 2. 
In the SMEFT, 
the operators $\tilde{\mathcal{O}}_a$ should be Lorentz scalars and singlets of the SM symmetry $SU(3)_C \times SU(2)_L \times U(1)_Y$. 
Their building blocks are the Higgs field $H$ (dimension 1), the covariant derivative $D$ (dimension 1),
the SM fermion $\psi$ (dimension 3/2), and the SM gauge field strength tensor $X$ (dimension 2). 
Due to the Lorentz symmetry, the only possibilities of the dimension-two operators $\tilde{\mathcal{O}}_a$
are thus two classes, $H^2$ and $D^2$. 
All independent operators of those classes have been listed in ref.~\cite{Grzadkowski:2010es}. 
The operators $\tilde{\mathcal{O}}_a$ are exhausted by $\tilde{\mathcal{O}}_{H} \equiv H^\dagger H$. 
Therefore, the dimension-two CS-type operator is given by
    \begin{equation}\label{dim6}
        \frac{\alpha}{8\pi} \frac{H^\dagger H}{\Lambda_H^2} \, F_{\mu\nu} \tilde{F}^{\mu\nu} \;.
    \end{equation} 
\\


{\bf Dimension 3 --} We next write down effective operators $\tilde{\mathcal{O}}_a$ of dimension 3. 
The Lorentz symmetry restricts the candidates to three classes such as $H^3$, $H D^2$ and $\psi^2$. 
However, the $SU(2)_L$ symmetry requires that the operators should contain an even number of $H$. 
This rules out all bosonic candidates. 
Since there is no hypercharge singlet of the SM fermion bilinear $\psi^2$, 
we can conclude that the SMEFT does not contain dimension-three CS-type operators \cite{Lehman:2014jma}. 
However, the operators of the $\psi^2$ type can arise in the LEFT respecting $SU(3)_C \times U(1)_{\rm EM}$. 
It has been shown in ref.~\cite{Liao:2020zyx} that
the CS-type operators of the $\psi^2$ type are generated by loop effects. 
In the LEFT, there are four-fermion interactions $(\bar{\psi} \Gamma \psi)(\bar{\psi}_c \Gamma \psi_c)~(\Gamma=1,\gamma_\mu,\sigma_{\mu\nu})$ with charged particles $\psi_c$. 
Among them, the scalar-type interaction $(\bar{\psi} \psi)(\bar{\psi}_c \psi_c)$ generates the CS-type operators. 
The Lagrangian (\ref{eq:L_CS}) is thus given by
    \begin{equation}
        \sum_{\psi=e, \nu, d, u}
        \frac{\alpha}{8\pi} \frac{\tilde{\mathcal{O}}_\psi}{\Lambda_\psi^3} F_{\mu\nu} \tilde{F}^{\mu\nu} \;,
        \label{dim7}
    \end{equation}
for three generations of charged leptons $e_i \, (i=1,2,3)$, neutrinos $\nu_i$ and
down-type quarks $d_i$, and two generations of up-type quarks $u_i$. 
Here, when the neutrinos are assumed to be Dirac fermions, 
the dimension-3 operators $\tilde{\mathcal{O}}_a$ are composed of the following fermion bilinears, 
\begin{subequations}
    \begin{align}
        \tilde{\mathcal{O}}_{e} &\equiv \tilde{\mathcal{C}}^{ij}_{e} \bar{e}{}^i P_{L} e^j  + {\rm h.c.}  \, ,
        \label{ebilinear} \\[1ex]
        \tilde{\mathcal{O}}_{\nu} &\equiv \tilde{\mathcal{C}}^{ij}_{\nu} \bar{\nu}{}^i P_{L} \nu^j  + {\rm h.c.}  \, , 
        \label{nubilinear} \\[1ex]
        \tilde{\mathcal{O}}_{d} &\equiv \tilde{\mathcal{C}}^{ij}_{d} \bar{d}{}^i P_{L} d^j  + {\rm h.c.}  \, , 
        \label{dbilinear} \\[1ex]
        \tilde{\mathcal{O}}_{u} &\equiv \tilde{\mathcal{C}}^{ij}_{u} \bar{u}{}^i P_{L} u^j  + {\rm h.c.}  \, ,
        \label{ubilinear}
    \end{align}
\end{subequations}
where $\tilde{\mathcal{C}}_{e, \nu, d, u}$ denote dimensionless coupling matrices,
$\bar{f} \equiv f^\dagger \gamma^0$ and $P_{L} \equiv (1-\gamma^5)/2$.
For Majorana neutrinos, we can define the similar operator $\tilde{\mathcal{O}}_{\nu}$ but
an extra factor of $1/2$ should be included for $i=j$.
In this case, the operator violates the lepton number by two units. 
When the interaction (\ref{dim7}) is matched to the SMEFT, 
the energy scale $\Lambda_\psi$ is related to a characteristic energy scale in the SMEFT, $\Lambda_{\rm SMEFT}$, as {\cite{Liao:2020zyx}}
    \begin{equation}
        \frac{1}{\Lambda_{\psi}^3} \sim \frac{v}{4\pi m_c}\frac{1}{\Lambda_{\rm SMEFT}^3} \,,
    \end{equation}
with the Higgs vacuum expectation value (VEV) $v$ and the mass $m_c$ of a charged particle in the loop. 

The vector-type interaction $(\bar{\psi} \gamma_\mu \psi)(\bar{\psi}_c \gamma^\mu \psi_c)$ such as the Fermi's interactions in the SM does not generate a CS-type operator. 
Instead, it could only generate an operator in the form:
    \begin{equation}\label{vector_CS}
        (\bar{\psi} \gamma_\mu \psi)K^\mu \;; ~ K^\mu \equiv A_\nu \tilde{F}^{\mu\nu} \,,
    \end{equation}
because of its tensor structure. 
As noted below Eq.~(\ref{op:jk}), 
the operator (\ref{vector_CS}) is not invariant under $U(1)_{\rm EM}$ unless $\bar{\psi} \gamma_\mu \psi = \nabla_\mu \tilde{\cal O}$ for a Lorentz scalar $\tilde{\cal O}$. 
However, this is not the case because the current $\bar{\psi} \gamma_\mu \psi$ has the transverse components. 
Therefore, we conclude that the operator (\ref{vector_CS}) is forbidden by $U(1)_{\rm EM}$ (see also ref.~\cite{El-Menoufi:2015dra}). 
\\


{\bf Dimension 4 --}
The operators such as $H^4$ and $H\psi^2$ in the SM Lagrangian can appear in effective operators $\tilde{O}_a$ of dimension 4. 
However, they are composed of the building blocks that have already appeared in the dimension-two/three operators.
Then, we only consider novel candidates,
    \begin{equation}
        \sum_{X=F, Z, W, G}
        \frac{\alpha}{8\pi} \left(\frac{X_{\alpha\beta}X^{\alpha\beta}}{\Lambda_X^4} + \frac{X_{\alpha\beta}\tilde{X}^{\alpha\beta}}{\Lambda_{\tilde X}^4}\right)F_{\mu\nu} \tilde{F}^{\mu\nu} \;,
        \label{dim8}
    \end{equation}
where $X, \tilde{X}$ denote a field-strength tensor and its dual,
and $Z,W,G$ are the $Z,W$ bosons and gluon, respectively.
On the CMB scale $T_{\rm LSS} \sim 0.3 \, {\rm eV}$, 
these types of operator with $X_{\mu\nu}=F_{\mu\nu}$ emerge from the electron loop \cite{Heisenberg:1936nmg,Schwinger:1951nm}: 
    \begin{equation}
        \mathcal{L}_{\rm EH} \supset \frac{7\alpha^2}{360 m_e^4}  \left( F_{\mu\nu} \tilde{F}^{\mu\nu}\right)^2 \;,
        \label{eq:EH}
    \end{equation}
with the electron mass $m_e$. This interaction is known as the Euler–Heisenberg Lagrangian,
which is one example of non-linear electrodynamics.
\\


{\bf Dimension $n \, (>4)$ --}
Much higher dimensional operators $\tilde{O}_a$ do not contain new building blocks and will give subdominant effects. 
Therefore, we do not consider such operators any further. 

\section{Isotropic Cosmic birefringence}
\label{ICB}

Let us discuss whether the listed CS-type operator of each dimension is able to induce the reported nonzero ICB with the corresponding cosmological background. 
The Lagrangian of interest is given by
    \begin{equation}
    \label{eq:LA}
        \mathcal{L}_A = -\frac{1}{4} F_{\mu\nu} F^{\mu\nu} - \frac{1}{4} \tilde{\cal O} F_{\mu\nu} \tilde{F}^{\mu\nu}\; .
    \end{equation}
On a cosmological background $\phi_{\tilde{\cal O}} \equiv \langle \tilde{\cal O} \rangle$, 
the CS-type operator $\tilde{\cal O} F_{\mu\nu} \tilde{F}^{\mu\nu}/4$ reduces to $-\phi_{\tilde{\cal O}}{\bf E} \cdot {\bf B}$. 
Then, the induced ICB angle $\beta$ is estimated as (see appendix \ref{a:macro} for the derivation)~\cite{Komatsu:2022nvu}
    \begin{align}\label{eq:beta}
        \beta =  \frac{1}{2} \int^{t_0}_{t_{\rm LSS}} dt \, \frac{\partial \phi_{\tilde{\cal O}}}{\partial t} 
         =  \frac{1}{2} \left[ \phi_{\tilde{\cal O}}(t_0) - \phi_{\tilde{\cal O}}(t_{\rm LSS}) \right]
        \, ,
    \end{align}
where $t_0, t_{\rm LSS}$ denote the present time and the time at the last scattering surface (LSS), respectively. 
Note that there can be the terms, $c_{EE} |{\bf E}|^2$ and $c_{BB} |{\bf B}|^2$ with time-dependent coefficients in the SMEFT/LEFT. 
We assume that the time dependence of these terms is tuned to be small not to contradict the constraints on the time variation of the fine structure constant \cite{Planck:2014ylh}. 
\\


{\bf Dimension 2 --}
First, we discuss the CS-type operator (\ref{dim6}),
    \begin{equation}
        \frac{\alpha}{8\pi} \frac{H^\dagger H}{\Lambda_H^2} \, F_{\mu\nu} \tilde{F}^{\mu\nu} \;.
    \end{equation} 
After the electroweak phase transition, the Higgs field gets a VEV $v$. 
We can neglect excitations from the vacuum because they are unstable and decay quickly. 
In the standard scenario, the VEV neither contributes to the ICB
because it leads to the CS-type operator with a time-independent coefficient. 
If the VEV $v$ depended on the time, 
it could induce the ICB. 
However, it would simultaneously induce  the time variation of the electron mass $m_e$ from the LSS to today, 
which is constrained to be $\Delta m_e/m_e = (4 \pm 11) \times 10^{-3} (68\%~{\rm C.L.})$\cite{Planck:2014ylh}. 
This means that the time variation of the VEV is at most a few percent of the electroweak scale. 
On the other hand, $\Lambda_H$ should be larger than the TeV scale from collider constraints \cite{Ellis:2020unq}. 
Therefore, we conclude that the operator (\ref{dim6}) cannot explain the reported nonzero ICB. 
\\


{\bf Dimension 3 --}
We next discuss whether each term of the CS-type operators \eqref{dim7},
    \begin{equation}
        \sum_{\psi=e, \nu, d, u}
        \frac{\alpha}{8\pi} \frac{\tilde{\mathcal{O}}_\psi}{\Lambda_\psi^3} F_{\mu\nu} \tilde{F}^{\mu\nu} \;,
    \end{equation}
with the fermion bilinears of Eqs.~\eqref{ebilinear}, \eqref{nubilinear}, \eqref{dbilinear}, \eqref{ubilinear},
can induce the reported ICB. 
The cosmic electron background is excluded because of the neutrality as we discussed in section \ref{s:relevant op}. 
The quark bilinear condensate after the QCD transition neither contributes to the ICB
because the condensate is independent of time and thus the resultant CS term becomes a total derivative. 
Therefore, 
the cosmic neutrino background (C$\nu$B) seems to be the most relevant to the ICB
and potentially able to explain the reported angle. 

The C$\nu$B is predicted to be generated from the thermal bath in the early Universe \cite{Weinberg:1962zza}. 
In the standard Big-Bang cosmology, its number density is comparable to that of the CMB photons. 
While the C$\nu$B has not been directly detected yet, its contribution to the radiation density has been detected by the WMAP 5-year observation \cite{WMAP:2008lyn} and later confirmed via the Planck observation \cite{Planck:2018vyg}.
Hence, the CS-type interaction \eqref{dim7} with the neutrino bilinear \eqref{nubilinear}
may induce the reported ICB during the photon propagation through the C$\nu$B.
We can rewrite the neutrino bilinear operator \eqref{nubilinear} as
    \begin{align}
        \tilde{\mathcal{O}}_{\nu}
        = \frac{(\tilde{\mathcal{C}}_{\nu}^\dagger+\tilde{\mathcal{C}}_{\nu} )^{ij}}{2} \bar{\nu}{}^i \nu^j
        + \frac{(\tilde{\mathcal{C}}^\dagger_{\nu} -\tilde{\mathcal{C}}_{\nu})^{ij}}{2} \bar{\nu}{}^i \gamma^5 \nu^j \, .
        \label{nubilinear2}
    \end{align}
Since we are now interested in the evolution of the photon field in the presence of the C$\nu$B,
the neutrino bilinear operator~\eqref{nubilinear2} is replaced with its background value
that is given by the expectation value $\langle \tilde{\mathcal{O}}_\nu \rangle$
with regard to a state of fixed neutrino and anti-neutrino number densities.
As calculated in appendix~\ref{backgroundnu},
we obtain $\langle \bar{\nu}{}^i \gamma^5 \nu^j \rangle = 0$ and
    \begin{align}
        \langle \bar{\nu}{}^i \nu^j \rangle & =  \delta^{ij} \mathcal{F}(t) \, , \label{eq:Dq1} \\[1ex]
        \mathcal{F}(t) &\equiv \int \frac{{d}^3 p}{\left( 2 \pi \right)^3} \frac{m_i}{E_{\mathbf{p}}}\left[n^i(p,t)+\bar{n}{}^i(p,t)\right] , \label{eq:f}
    \end{align}
where $n^i, \bar{n}^i$ denote the phase-space number densities of the $i$-th neutrino and anti-neutrino, respectively,
and $m_i$ is the neutrino mass.
As noted at the beginning of Sec.~\ref{s:relevant op}, we neglect small effects from the cosmic expansion (as well as those from the spacetime curvature) in the present calculations.
Thus, the function $\phi_{\tilde{\cal O}}$ in Eq.~(\ref{eq:beta}) is given by
    \begin{align}
        \phi_{\tilde{\cal O}}(t) =
        \frac{\alpha}{4\pi} \frac{{\rm tr} [ ( \tilde{\mathcal{C}}_{\nu} +\tilde{\mathcal{C}}{}^\dagger_{\nu} ) \mathcal{F}(t) ]}{\Lambda_{\nu}^3} \, ,
    \end{align}
and we put an extra factor $1/2$ for the case of Majorana neutrinos as noted below Eq.~(\ref{ubilinear}). 
Since $\phi_{\tilde{\cal O}}$ redshifts due to the cosmic expansion, 
we can well approximate the ICB angle as $\beta \simeq - \phi_{\tilde{\cal O}}(t_{\rm LSS}) / 2$. 
At the time of the last scattering, the temperature of the Universe is $T_{\rm LSS} \sim 0.3 \, {\rm eV}$.
Assuming $m_i \ll T_{\rm LSS}$ at the last scattering for some or all of the neutrino species, we can analytically evaluate $\mathcal{F}$ in Eq.~\eqref{eq:f}, 
\begin{align}
\mathcal{F}(t_{\rm LSS}) \simeq 
0.5 \,
\frac{m_i}{T_{\rm LSS}} \left( N^i + \bar{N}^i \right)  , \, \, \,
m_i \ll T_{\rm LSS} \, ,
\end{align}
where $N^i$ and $\bar{N}^i$ are the number densities of the neutrino and anti-neutrino, respectively, at the LSS. 
Therefore, we obtain
    \begin{align}\label{beta nu}
        \beta \simeq
        - 0.008^{\, \circ} \frac{\alpha}{137^{-1}}
        \sum_i \frac{m_i}{T_{\rm LSS}} (  \tilde{\mathcal{C}}_{\nu} +\tilde{\mathcal{C}}{}^\dagger_{\nu}  )^{ii} \frac{N^i + \bar{N}^i}{\Lambda_\nu^3} \; .
    \end{align}
Here, the neutrino number density at the last scattering is estimated to be $N_i^{1/3} = {\cal O}(10^{-10})\,{\rm GeV}$ in the natural unit. 
The CS-type neutrino interaction is constrained by various experiments measuring coherent elastic neutrino-nucleus scattering,
deep inelastic neutrino–nucleon scattering
and solar neutrino scattering,
as well as collider searches~\cite{Altmannshofer:2018xyo}.
The resulting lower bound on the mass scale $\Lambda_\nu$ ranges from
$\mathcal{O}(10^{-2}) \, {\rm GeV}$ to $\mathcal{O}(10^{2}) \, {\rm GeV}$.
Then, we find that the ICB angle~\eqref{beta nu} would be much smaller than the observed value. 
\\


{\bf Dimension 4 --}
Finally, we discuss the CS-type operator (\ref{dim8}). 
To see the propagation of a photon in the cosmological magnetic field, we separate the field strength into the background and propagating-photon parts $F_{\mu\nu} = F_{\mu\nu}^{\rm (bg)} + F_{\mu\nu}^{\rm (p)}$. 
When the background $F_{\mu\nu}^{\rm (bg)}$ is pure magnetic field, 
the operators (\ref{dim8}) give
    \begin{equation}
        (F_{\alpha\beta}^{\rm (bg)}F^{{\rm (p)}\alpha\beta})(F^{\rm (bg)}_{\mu\nu}\tilde{F}^{{\rm (p)}\mu\nu})\;, ~
        (F^{\rm (bg)}_{\mu\nu}\tilde{F}^{{\rm (p)}\mu\nu})^2\;,
        \label{eq:quadratic_FF}
    \end{equation}
as well as the CS-type term
    \begin{equation}
        (F_{\alpha\beta}^{\rm (bg)}F^{{\rm (bg)}\alpha\beta})(F^{\rm (p)}_{\mu\nu}\tilde{F}^{{\rm (p)}\mu\nu})\;,
        \label{eq:quadratic_FF_CS}
    \end{equation}
to the quadratic action of the propagating-photon field. 
Here, their coefficients are of the same order. 
The operator (\ref{eq:quadratic_FF_CS}) gives 
the CS-type term ${\bf E} \cdot {\bf B}$. 
Meanwhile, the operators (\ref{eq:quadratic_FF}) reduce to ${\bf E}_\parallel \cdot {\bf B}_\parallel$ in Eq.~(\ref{op:aniso pv}) and ${\bf E}_\parallel \cdot {\bf E}_\parallel$ in Eq.~(\ref{op:aniso pc}). 
Therefore, the operators (\ref{dim8}) inevitably cause unwanted ACB as well as the ICB (see the last paragraph in section \ref{s:relevant op}). 

As for the gauge fields $X=Z, W, G$, the weak gauge bosons are excluded because they are unstable and decay quickly. 
The gluon forms a nonzero condensate and contributes to the ICB. 
However, it would be negligibly small because the energy scale of the condensate is the QCD scale while the scale $\Lambda_X$ is constrained from collider experiments as $\Lambda_X \gtrsim 1 \, {\rm TeV}$ \cite{Ellis:2018cos}. 
It is also notable that there are other operators with the same symmetry, {\it e.g.,} $(X_{\alpha\beta}F^{\alpha\beta}) (X_{\mu\nu}\tilde{F}^{\mu\nu})$, which would induce the ACB and need to be suppressed. 

\section{Beyond SMEFT/LEFT}
\label{s:extensions}

We can extend our arguments to narrow down possible new particles that are able to explain the reported nonzero ICB. 
Unless a particle is a SM singlet like the ALP, 
the leading CS-type operator is given by
    \begin{align}\label{CS dark}
        \begin{aligned}
        &\frac{\alpha}{8\pi}\frac{\Phi^\dagger \Phi}{\Lambda^2}F_{\mu\nu}\tilde{F}^{\mu\nu} \quad
        \text{(for a scalar $\Phi$)}
        \;, \\[1ex]
        &\frac{\alpha}{8\pi}\frac{\bar{\chi} \chi}{\Lambda^3}F_{\mu\nu}\tilde{F}^{\mu\nu} \quad
        \text{(for a fermion $\chi$)} 
        \;.
        \end{aligned}
    \end{align}
For the operator to induce the ICB, $\Phi^\dagger \Phi$ or $\bar{\chi} \chi$ should have a time-dependent background value. 
There would be three possibilities: the cosmological background of $\Phi^\dagger \Phi$ or $\bar{\chi} \chi$
is composed of (i) classical fields, (ii) pair condensates, or (iii) particles. 
The case (i) is similar to the ALP case. 
In the case (ii), the pair condensates would also effectively work as axion-like fields. 
To discuss this possibility, 
we need to elaborate a model with interaction to form appropriate time-dependent condensates
and its cosmological consequences, which is left for future studies. 
In the following, we focus on the possibility (iii).


Since the one-particle energy $E_{\mathbf{p}}$ is always larger than the mass $m$, 
the cosmological background of $\Phi^\dagger \Phi$ or $\bar{\chi}\chi$ is bounded from above by the energy density $\rho$ as
    \begin{align}
        \langle \Phi^\dagger \Phi \rangle
        \lesssim \rho/m^2 \;, \quad 
        \langle \bar{\chi}\chi \rangle
        \lesssim \rho/m \;,
    \end{align}
respectively. 
In the epoch after the LSS, the energy density is conservatively bounded by the critical density at the LSS, $\rho_{\rm c, LSS} \simeq (3 \times 10^{-13} {\rm TeV})^4$ (natural unit). 
Substituting these results to Eq.~\eqref{eq:beta}, 
we find
    \begin{align}\label{eq:limit on ds}
        \begin{aligned}
            m & 
            \lesssim 10^{-14} \, {\rm eV}\left(\frac{|\beta|}{0.3^{\, \circ}}\right)^{-1/2}\left( \frac{\Lambda}{\rm TeV} \right)^{-1} ~ \text{(scalar)} \;, \\[1ex]
            m & 
            \lesssim 10^{-40}\, {\rm eV} \left(\frac{|\beta|}{0.3^{\, \circ}}\right)^{-1}\left(\frac{\Lambda}{\rm TeV}\right)^{-3} ~ \text{(fermion)}\;.
        \end{aligned}
    \end{align}
The interactions (\ref{CS dark}) can be probed by collider experiments through production processes, $\gamma \to \gamma \Phi \Phi$, $\gamma \chi \chi$. 
As shown for the fermion $\chi$ in ref.~\cite{ATLAS:2017nga}, 
the energy scale $\Lambda$ should be roughly larger than the TeV scale due to the absence of such a process
provided that it is kinematically allowed, {\it i.e.,} the mass is smaller than the TeV scale. 
Then, the conditions (\ref{eq:limit on ds})
indicate that the mass of the particle should be extremely small. 
In addition to potential missing energy carried by such light particles in electromagnetic signals, this scenario has the same theoretical problem as the ALP: 
the existence of such ultra-light particles with the CS-type interactions (\ref{CS dark}) has tension with simple Grand Unified models because interactions with gluons give a large contribution to the mass.
It is also noted that the particles should never be thermalized. 
Otherwise, we could apply the argument similar to the neutrino case, and thus the induced ICB angle would become negligibly small. 
This requirement restricts possible interactions with the SM particles. 
Moreover, the maximum temperature of the Universe, denoted as $T_{\rm max}$, has an upper limit so that the particles are never thermalized through the interactions (\ref{CS dark}). 
Roughly estimating the interaction rate as $\Gamma \sim (\alpha/8\pi)^2 T^5/\Lambda^4 ~ (\text{scalar})$ or $(\alpha/8\pi)^2 T^7/\Lambda^6~ (\text{fermion})$, we can find the upper limit from the Gamow's criterion $\Gamma < H$ as
    \begin{align}\label{eq:Tmax}
        \begin{aligned}
            T_{\rm max} &\lesssim 1 \,{\rm MeV} \left(\frac{\Lambda}{3 \,{\rm GeV}}\right)^{4/3} ~ \text{(scalar)} \;, \\
            T_{\rm max} &\lesssim 1 \, {\rm MeV} \left(\frac{\Lambda}{0.2 \, {\rm GeV}}\right)^{6/5} ~ \text{(fermion)}\;.
        \end{aligned}
    \end{align}
This argument implies that the energy scale $\Lambda$ should be roughly larger than the GeV scale 
because $T_{\rm max}$ must be larger than the temperature of the Big Bang Nucleosynthesis,
$T_{\rm BBN} \sim 1\, {\rm MeV}$. 


A similar argument can be applied to a dark vector field $V_\mu$ such as a dark photon. 
In the massless case, the leading CS-type operator is given by Eq.~(\ref{dim8}) with the field strength $X_{\alpha\beta}$ for the vector field $V_\mu$. 
The ICB angle induced by this operator is much suppressed because the field strength is bounded by the energy density as $|X_{\alpha \beta}X^{\alpha \beta}|<4\rho$. 
In the massive case, the CS-type operator $(V_\alpha V^\alpha) F_{\mu\nu}\tilde{F}^{\mu\nu}$ is also allowed. 
When this operator gives the leading contribution to the ICB angle, the argument is parallel to the scalar case: the mass should be extremely small. 


Another possibility is that the $U(1)_{\rm EM}$ symmetry is only realized through a Stueckelberg field so that 
the operator (\ref{op:jk}) is allowed. 
As mentioned above, this operator is reduced to $c_{\rm EB} {\bf E} \cdot {\bf B}$ with $\dot{c}_{\rm EB} = J_0$. 
Roughly estimating $\dot{c}_{\rm EB} \sim H c_{\rm EB}$ with the Hubble parameter $H$, 
we see that the induced ICB angle $\beta$ could be comparable to the reported value $\beta \sim {\cal O} (0.1^{\, \circ})$ for the neutrino background $J_0 \sim n_\nu$ due to the significant enhancement factor $H^{-1}$~\cite{Geng:2007va, Ho:2010aq, Mohammadi:2021xoh, Zhou:2023aqz}.
However, in this scenario,  
the symmetry allows the photon mass in the vacuum. 
For example, a Stueckelberg scalar makes it possible to add the photon mass term $A_\mu A^\mu$ in the action. When a Stueckelberg field is given by a two-form field $B_{\mu\nu}$~\cite{Geng:2007va, Ho:2010aq, Mohammadi:2021xoh, Zhou:2023aqz}, we can add the BF term $B_{\mu\nu} \tilde{F}^{\mu\nu}$, which is known to induce the photon mass \cite{Cremmer:1973mg}. Since the experimental upper bound on the photon mass is ${\cal O}(10^{-18})\,{\rm eV}$~\cite{Goldhaber:2008xy, ParticleDataGroup:2022pth}, 
we need a mechanism to forbid or suppress these operators.

\section{Conclusion}
\label{conclusion}

In the present paper, we have investigated the interpretation of the reported nonzero ICB angle.
Adopting the EFT approach, it is concluded that (1) only a CS-type operator could produce such a parity-violating ICB effect in the presence of a cosmic background that is assumed to be homogeneous and neutral,
and (2) no SM particles can explain the reported value under the standard cosmological evolution.
Among all SM contributions, the C$\nu$B could be the most promising, but its contribution has been calculated explicitly and found to be too small to explain the reported ICB.
Our result would indicate the existence of a new particle lighter than the electroweak scale or some exotic cosmological scenarios if the reported value of ICB should be confirmed.
We have provided a guideline on searching for physics beyond the SM through the ICB apart from the ALP.
Some constraints on a dark sector particle and the maximum temperature of the Universe have been discussed
by assuming that the dark sector is responsible for the ICB, which might be helpful in identifying the dark matter.

%
\section*{Acknowledgments}
We would like to thank Eiichiro Komatsu and Satoshi Shirai for discussions. 
This work is supported by Natural Science Foundation of China No.~12150610465 (YN), the RIKEN Incentive Research Project grant (RN), and JSPS Overseas Research Fellowship / JSPS KAKENHI No.~JP20H05859 and 19K14702 (IO), No.~19H01891, No.~20H05860 (RS).

\appendix

\section{Derivation of the ICB angle $\beta$}
\label{a:macro}

In this appendix, we give a derivation of Eq.~(\ref{eq:beta}), given \eqref{eq:LA}, for the ICB angle $\beta$ induced by a CS-type operator,
    \begin{equation}
        -\frac{\phi_{\tilde{\cal O}}}{4}F_{\mu\nu}\tilde{F}^{\mu\nu} = \phi_{\tilde{\cal O}} {\bf E} \cdot {\bf B} \;,
        \label{eq:cs-term}
    \end{equation}
following the discussion in ref.~\cite{Nieves:1988qz}. 
We assume that a photon propagates in the homogeneous and isotropic background,
$ds^2 = a(\eta)(- d\eta^2 + d\vec{x}^2)$, where $a(\eta)$ is the scale factor and $\eta$ is the conformal time.
Since it can be conformally transformed into the Minkowski spacetime,
we evaluate every quantity on the Minkowski spacetime in this appendix.

In the presence of the CS-type operator (\ref{eq:cs-term}), 
Maxwell equations are modified. 
We first write down general Maxwell equations for the spatially-invariant operators (\ref{op:EFT spatial}).
In the momentum space, ${\bf E}(x) \to \tilde{\bf E}(k) e^{i k \cdot x}$ and ${\bf B}(x) \to \tilde{\bf B}(k) e^{i k \cdot x}$,%
\footnote{This $4$-vector Fourier transformation is justified for slow variations of the source terms compared to the oscillation time scale due to $\omega$ and ${\bf k}$, which is consistent with our underlying assumption in considering ICB.}
they can be written as
    \begin{subequations}
        \begin{align}
        i {\bf k} \cdot \tilde{\bf E} & =  \rho_{\rm ext} + \rho_{\rm ind} \ , \label{eq:inhomo1} \\ 
        {\bf k} \cdot \tilde{\bf B} & = 0 \ , \label{eq:homo1}\\
        {\bf k} \times \tilde{\bf E} & = - \omega \tilde{\bf B} \ ,  \label{eq:homo2}\\ 
        i {\bf k} \times \tilde{\bf B} & =  {\bf j}_{\rm ext} + {\bf j}_{\rm ind} + i \omega \tilde{\bf E}\;,\label{eq:inhomo2}
        \end{align}
    \end{subequations}
where $\omega \equiv k_0$. 
In addition to the external sources $\rho_{\rm ext}$ and $\mathbf{j}_{\rm ext}$, 
the source terms $\rho_{\rm ind} = \rho_{\rm ind}(\tilde{\bf E},\tilde{\bf B})$ and ${\bf j}_{\rm ind} = {\bf j}_{\rm ind}(\tilde{\bf E},\tilde{\bf B})$ are induced by the operators (\ref{op:EFT spatial}) other than the standard one. 
These terms can be rewritten only in terms of the electric field $\tilde{\bf E} $ by using the Maxwell equation \eqref{eq:homo2}: $\rho_{\rm ind} = \rho_{\rm ind}(\tilde{\bf E})$ and ${\bf j}_{\rm ind} = {\bf j}_{\rm ind}(\tilde{\bf E})$.
From Eq.~\eqref{eq:homo2} and Eq.~\eqref{eq:inhomo2}, we have
    \begin{equation}
        {\bf j}_{\rm ext} + {\bf j}_{\rm ind} = - i \omega \left\{\tilde{\bf E}- \frac{|{\bf k}|^2}{\omega^2} \left[\tilde{\bf E}- \hat{\bf k}(\hat{\bf k}\cdot\tilde{\bf E}) \right]\right\} \;,
        \label{eq:j_ext+j_ind}
    \end{equation}
where $\hat{\bf k} \equiv {\bf k}/|{\bf k}|$. 
Decomposing $\tilde{\bf E}$ into the components parallel and transverse to the wave vector ${\bf k}$ as
    \begin{equation}
    \label{eq:def_ElEt}
        \tilde{\bf E}_{\rm l} = \hat{\bf k} ( \hat{\bf k}\cdot \tilde{\bf E})\;, \quad \tilde{\bf E}_{\rm t} = \tilde{\bf E} - \tilde{\bf E}_{\rm l}\;,
    \end{equation}
we can rewrite Eq.~(\ref{eq:j_ext+j_ind}) as
    \begin{equation}
        {\bf j}_{\rm ext} + {\bf j}_{\rm ind} = - i \omega \left[ \tilde{\bf E}_{\rm l} + \left(1 - \frac{|{\bf k}|^2}{\omega^2}\right) \tilde{\bf E}_{\rm t} \right] \;.
         \label{eq:j_ext+j_ind El Et}
    \end{equation}
Since the operators in \eqref{op:EFT spatial} are quadratic in the electromagnetic fields, 
the induced current ${\bf j}_{\rm ind} = {\bf j}_{\rm ind}(\tilde{\bf E})$ is linear in $\tilde{\bf E}$. 
The three spatial vectors $\tilde{\bf E}_{\rm t}$, $\tilde{\bf E}_{\rm l}$, and $\hat{\bf k}\times\tilde{\bf E} = \hat{\bf k} \times \tilde{\bf E}_{\rm t}$ span the three-dimensional space, provided that neither $\tilde{\bf E}_{\rm l}$ nor $\tilde{\bf E}_{\rm t}$ is null. 
Thus, 
in general, we can parameterize the induced current as
    \begin{equation}
        \mathbf{j}_{\rm ind} = - i\omega\left[ (1- \varepsilon_{\rm l}) \tilde{\bf E}_{\rm l} + \left( 1- \varepsilon_{\rm t} \right) \tilde{\bf E}_{\rm t} - i \varepsilon_{\rm p} \hat{\mathbf{k}} \times \tilde{\bf E}  \right]\;.
        \label{eq:j_ind}
    \end{equation}
Here, the three parameters $\varepsilon_{\rm l}$, $\varepsilon_{\rm t}$, and $\varepsilon_{\rm p}$ consist of $\omega$ and
three arbitrary functions $c_{\rm EE}$, $c_{\rm BB}$, and $c_{\rm EB}$ in Eq.~(\ref{op:EFT spatial}). 
As understood from the parity, the $\varepsilon_{\rm p}$ term is generated by the CS-type operator (\ref{eq:cs-term}). 
Taking into account the assumption that $\phi_{\tilde{\mathcal{O}}}$ is a function only of the time $\eta$, the induced current from the CS-type operator (\ref{eq:cs-term}) is given by
    \begin{align}
        {\bf j}_{\rm ind} = \phi'_{\tilde{\cal O}}(\eta) \tilde{\bf B} = - \phi'_{\tilde{\cal O}}(\eta) \, \frac{{\bf k} \times \tilde{\bf E}}{\omega} \,,
    \end{align}
where a prime denotes a derivative with respect to $\eta$,
and $\varepsilon_{\rm p}$ can be read off as
    \begin{equation}
        \varepsilon_{\rm p} = 
        \phi'_{\tilde{\cal O}}(\eta) \, \frac{|{\bf k}|}{\omega^2} \;.
        \label{eq:cs ep_p}
    \end{equation}
Here, we have assumed that $\phi'_{\tilde{\cal O}}$ is approximately constant, {\it i.e.,} $|\phi'_{\tilde{\cal O}}/\phi_{\tilde{\cal O}}| \ll \omega, |{\bf k}|$, which is valid for the CMB photons propagating in the cosmological background. 
We now consider an electromagnetic (EM) wave propagating through media only with the induced charge current, $\mathbf{j}_{\rm ext} = 0$. 
Let us define
    \begin{equation}
        \mathbf{E}_{\pm} = \frac{1}{2} \left( \tilde{\bf E}_{\rm t} \pm i \hat{\mathbf{k}} \times \tilde{\bf E} \right) \;.
        \label{eq:pm_mode}
    \end{equation}
In the absence of external source, i.e.~$\rho_{\rm ext} = 0$, with $\phi_{\tilde{\cal O}}$ having only time dependence, which gives $\rho_{\rm ind} = 0$, Eqs.~\eqref{eq:inhomo1} and \eqref{eq:def_ElEt} tell us $\tilde{\bf E}_{\rm l} = 0$, which we set hereafter.%
\footnote{In this case, we would in principle have to replace $\tilde{\bf E}_{\rm l}$ in \eqref{eq:j_ind} by a term proportional to $\hat{k}$ in order to span the $3$-D space; however, one can trivially see that this term should be zero.}
Then, we can rewrite the wave equation (\ref{eq:j_ext+j_ind El Et}) by inserting Eq.~(\ref{eq:j_ind}) as
    \begin{align}
        0 & = i \omega \bigg[ 
        \left( \varepsilon_{\rm t} +\varepsilon_{\rm p} - \frac{|{\bf k}|^2 } {\omega^2} \right) \tilde{\bf E}_{+} 
        + \left( \varepsilon_{\rm t} - \varepsilon_{\rm p} - \frac{|{\bf k}|^2 } {\omega^2} \right) \tilde{\bf E}_{-}  \bigg]\;,
    \end{align}
and the following dispersion relations are satisfied:
    \begin{equation}
        \varepsilon_{\rm t} \pm \varepsilon_{\rm p} = |{\bf k}|^2/\omega^2\;,
        \label{eq:dispersion}
    \end{equation}
where $\pm$ denotes the $2$ transverse helicity modes.

As we mention in the main text, 
we only consider the CS-like operator (\ref{eq:cs-term}). 
Then, we focus on the case with $\varepsilon_{\rm t}=1$ and $\varepsilon_{\rm p}= \phi'_{\tilde{\cal O}}(\eta)|{\bf k}|/\omega^2$ [see Eq.~(\ref{eq:cs ep_p})]. 
The dispersion relations (\ref{eq:dispersion}) become
    \begin{equation}
        \omega_{\pm} = |{\bf k}| \sqrt{1 \mp \frac{\phi'_{\tilde{\cal O}}}{|{\bf k}|}} \approx |{\bf k}| \mp \frac{1}{2} \phi_{\tilde{\mathcal{O}}}'\;,
    \end{equation}
for the $\pm$-helicity modes, 
where we have 
assumed that $|\phi_{\tilde{\mathcal{O}}}'|\ll |{\bf k}|$ in the second equation. 
The phases for the $\pm$-helicity modes are estimated as
    \begin{align}
        \theta_\pm = \bar{\theta} \pm \beta \,; \quad \beta \equiv \frac{1}{2}\int \phi'_{\tilde{{\cal O}}}(\eta) {\rm d}\eta \,,
        \label{eq:theta_pm}
    \end{align}
where $\bar{\theta} \equiv |{\bf k}|(-\eta + \hat{\bf k} \cdot {\bf x})$.
Thus, from Eq.~(\ref{eq:pm_mode}), the monochromatic EM waves at LSS and today are related as
    \begin{align}
        {\bf E}(\eta_0) 
        &= {\bf E}_+(\eta_{\rm LSS})e^{i\theta_+} +  {\bf E}_-(\eta_{\rm LSS})e^{i\theta_-} \nonumber \\
        &=  e^{i\bar{\theta}}\left[ {\bf E}(\eta_{\rm LSS})\cos\beta - \hat{\bf k} \times {\bf E}(\eta_{\rm LSS}) \sin\beta \right] \;,
    \end{align}
where the phases are defined by taking the integration from $\eta_{\rm LSS}$ to $\eta_0$ in Eq.~(\ref{eq:theta_pm}). 
This equation shows that the polarization direction rotates clockwise by the angle $\beta$ with respect to the line-of-sight direction ${\bf n}_{\rm LOS} \equiv -\hat{\bf k}$. Therefore, we can conclude that the ICB angle is given by
    \begin{align}
    \beta
    = \frac{1}{2}\int_{\eta_{\rm LSS}}^{\eta_0} \phi_{\tilde{\mathcal{O}}}'(\eta) d\eta = \frac{1}{2}\left[\phi_{\tilde{\mathcal{O}}}(t_0)-\phi_{\tilde{\mathcal{O}}}(t_{\rm LSS})\right] \;,
    \end{align}
where we have replaced the conformal time with the cosmic time after the integration. 
Therefore, we have derived Eq.~(\ref{eq:beta}). 
Note that the birefringence effect is only generated by $\varepsilon_{\rm p}$, 
which is only induced by a CS-type operator.

\section{Polarization tensors in medium}
\label{a:propagation medium}

Let us revisit the argument in section \ref{s:relevant op} from the viewpoint of the photon
propagation in a medium, extending the analysis presented in ref.~\cite{Nieves:1988qz}. 
We explicitly show that the operator ${\bf E} \cdot {\bf B}$ (${\bf E}_\parallel \cdot {\bf B}_\parallel$)
generates the isotropic (anisotropic) cosmic birefringence. 
The effective kinetic action of the photon field $A_\mu$ is given by
    \begin{equation}
        S_{\rm kin} = \frac{1}{2} \iint {\rm d}^4 x \, {\rm d}^4 y \left[ A_\mu(x) (D^{-1})^{\mu\nu}(x,y) A_\nu(y) \right] \;,
        \label{eq:S_kin}
    \end{equation}
where $D_{\mu\nu}$ denotes the propagator in the presence of a cosmological background field $J$:
    \begin{equation}
        \langle TA_\mu(x)A_\nu(y) \rangle_J = iD_{\mu\nu}(x,y) \;,
    \end{equation}
for the time ordering $T$. 
Introducing the self energy $\Pi^{\mu\nu}$, 
we can write $(D^{-1})^{\mu\nu}$ as 
    \begin{equation}
        (D^{-1})^{\mu\nu} = (\Delta^{-1})^{\mu\nu} + \Pi^{\mu\nu} \;,
    \end{equation}
where $(\Delta^{-1})^{\mu\nu}$ is the tree-level part,
    \begin{equation}
        (\Delta^{-1})^{\mu\nu} \equiv -i(g^{\mu\nu}\nabla^2 -\nabla^\mu\nabla^\nu)\delta^4(x-y) \;.
    \end{equation}
In the Fourier space, 
the self-energy term in the kinetic action (\ref{eq:S_kin}) is written as
    \begin{equation}
        \frac{1}{2} \iint {\rm d}^4 k_1 {\rm d}^4 k_2 \left[ A_\mu(k_1) \Pi^{\mu\nu}(k_1,k_2)A_\nu(k_2) \right] \;,
    \end{equation}
and its gauge invariance requires that $\Pi^{\mu\nu}$ should satisfy
    \begin{equation}\label{gauge_con}
        k_1^{\mu} \Pi_{\mu\nu} = 0 \,, \quad k_2^{\nu} \Pi_{\mu\nu} = 0 \,. 
    \end{equation}
Moreover, 
    \begin{equation}\label{bose_con}
        \Pi^{\mu\nu}(k_1,k_2) = \Pi^{\nu\mu}(k_2,k_1) \,,
    \end{equation}
from the Bose-Einstein statistics and
    \begin{equation}\label{Lorentz}
        L^{\mu}{}_{\alpha}L^{\nu}{}_{\beta}\Pi^{\alpha\beta}(k_1,k_2; J) = \Pi^{\mu\nu}(L\cdot k_1, L\cdot k_2; L\cdot J) \,,
    \end{equation}
for a Lorentz transformation $L$. 
Here, $(L \cdot k_1)^\mu \equiv L^{\mu}{}_{\alpha} k_1^{\alpha}$ and so on. 
In the condition (\ref{Lorentz}), 
we have explicitly written the dependence on the background field $J$, 
which is not necessarily invariant under the Lorentz transformation.

In the vacuum without background, 
the four-momentum of the photon is conserved: $k_1^\mu + k_2^\mu = 0$. 
Thus, the independent tensors that can appear in $\Pi^{\mu\nu}$ are $g^{\mu\nu}$, $\epsilon^{\mu\nu\alpha\beta}$ and $k_1^{\mu}~(=-k_2^{\mu})$. 
The tensor structure of $\Pi^{\mu\nu}$ can be extracted as
    \begin{equation}
        \Pi^{\mu\nu}(k_1,k_2) \stackrel{\text{vac}}{=} |k_1|^2 \Pi(|k_1|^2) {\cal P}_1{}^{\mu\nu} \delta^4(k_1-k_2) \;,
    \end{equation}
where we have introduced the projection operator
    \begin{equation}
        {\cal P}_a{}^{\mu\nu} \equiv g^{\mu\nu} - \frac{k_a^\mu k_a^\nu}{|k_a|^2} \quad (a=1,2)\,.
    \end{equation}
This shows that the structure of the kinetic term is kept, ${\cal L}_{\rm kin} \sim F_{\mu\nu}F^{\mu\nu}$. 
That is, the ICB is not generated in the vacuum.

In the presence of a cosmological background, 
a background tensor $J$ should be added to the building blocks. 
The cosmological principle requires that $J$ is a function of the cosmic time $t_c$. 
Any tensor indices of $J$ and its derivatives should be determined by the unit-four vector $u^\mu \propto \nabla^\mu t_c$ as well as $g^{\mu\nu}$ and  $\epsilon^{\mu\nu\alpha\beta}$. 
Moreover, the four-momentum conservation is violated by the cosmological background as 
$k_1^\mu+k_2^\mu \propto u^\mu$. 
Then, the building blocks of $\Pi^{\mu\nu}$ are $g^{\mu\nu}$, $\epsilon^{\mu\nu\alpha\beta}$, $k_1^{\mu}$ and $u^\mu$. 
The possible tensor structures of $\Pi^{\mu\nu}$ with the properties (\ref{gauge_con})-(\ref{Lorentz}) are
    \begin{equation}\label{pi_medium}
        {\cal P}_1{}^{\mu}{}_{\alpha}{\cal P}_2{}^{\alpha\nu} \,, \quad ({\cal P}_1 \cdot k_2)^\mu ({\cal P}_2 \cdot k_1)^\nu \,, \quad \epsilon^{\mu\alpha\nu\beta}k_{1\alpha}k_{2\beta} \,,
   \end{equation}
with $k_1^\mu+k_2^\mu \propto u^\mu$, 
where $({\cal P}_a \cdot k_b)^\mu \equiv {\cal P}_a{}^{\mu\nu}k_{b\nu}~(a,b=1,2)$. 
These three tensors give all independent components in $\Pi^{\mu\nu}$. 
We can see this fact from the viewpoint of the photon propagation in the medium, 
which illuminates the role of the homogeneity and isotropy. 
As we can see from $k_1^\mu+k_2^\mu \propto u^\mu$, 
the homogeneity ensures that the momentum ${\bf k}$ is conserved in the CMB rest frame. 
If we introduce the helicity basis with respect to ${\bf k}$,
    \begin{equation}\label{helicity_basis}
        \left\{ \epsilon_L{}^\mu({\bf k}) \,, \epsilon_+{}^\mu({\bf k}) \,, \epsilon_-{}^\mu({\bf k}) \right \} 
        \quad ( \epsilon_a{}^\mu u_\mu = 0 ) \,,
    \end{equation}
the isotropy ensures that they are not mixed with each other in the propagation. 
Here, $\epsilon_L{}^\mu$ and $\epsilon_\pm{}^\mu$ respectively represent the longitudinal mode and the transverse modes with the helicity $\sigma = \pm$ in the sense that $k_\mu \epsilon_\pm^\mu = 0 \ne k_\mu \epsilon_L^\mu$.
Taking also into account that the temporal components are determined by the gauge condition (\ref{gauge_con}), 
we see that the independent components are three diagonal elements for the helicity basis (\ref{helicity_basis}):
    \begin{equation}
        \Pi^{\mu\nu} = \sum_{a=L,\pm} \Pi_a {\epsilon^\ast}_a{}^{\mu}\epsilon_a{}^{\nu} + \text{(temporal components)} \,,
    \end{equation}
where ${\epsilon^\ast}_a{}^{\mu}$ is the complex conjugate of $\epsilon_a{}^{\mu}$ (${\epsilon^\ast}_\pm{}^{\mu} = \epsilon_\mp{}^{\mu}$) 
and $\Pi_a$ is a scalar quantity constructed from $k_1^\mu$, $k_2^\mu$ and $u^\mu$. 
Therefore, the tensor structure of $\Pi^{\mu\nu}$ can be determined by three independent tensors.

It is straightforward to show that the tensor structure of the operator (\ref{op:EFT spatial})
is written as a linear combination of the three tensors (\ref{pi_medium})
by using the covariant form of ${\bf E}$ and ${\bf B}$,
    \begin{align}
        E^{\mu} = u_{\nu}F^{\mu\nu} \,, \quad B^{\mu} = u_{\nu} \tilde{F}^{\mu\nu} \,.
    \end{align}
As we can understand from the parity, 
the first two tensors in \eqref{pi_medium} correspond to $|{\bf E}|^2$ and $|{\bf B}|^2$. 
For example, 
the tensor structure of the operator $|{\bf E}|^2 - |{\bf B}|^2 = F_{\mu\nu}F^{\mu\nu}/2$ is given by
    \begin{align}
        \left[ (k_1 \cdot k_2){\cal P}_1{}^{\mu}{}_{\alpha}{\cal P}_2{}^{\alpha\nu} - ({\cal P}_1 \cdot k_2)^\mu ({\cal P}_2 \cdot k_1)^\nu \right] A_\mu A_\nu \,.
    \end{align}
The third tensor in \eqref{pi_medium} corresponds to ${\bf E} \cdot {\bf B} \propto F_{\mu\nu}\tilde{F}^{\mu\nu}$: 
    \begin{align}\label{int_kk}
        \epsilon^{\mu\alpha\nu\beta}k_{1\alpha}k_{2\beta}A_\mu(k_1)A_\nu(k_2)
        \propto F_{\mu\nu}(k_1)\tilde{F}^{\mu\nu}(k_2) \,.
    \end{align}
Therefore, any parity-violating operator should be given by the CS-type operator in the cosmological background.
Now, we see that the term $c_{\rm EB} {\bf E} \cdot {\bf B}$ in (\ref{op:EFT spatial}) generates the ICB.  
Using $k_1^\mu+k_2^\mu \propto u^\mu$, 
we can write the four-momenta $k_a^\mu~(a=1,2)$ as
    \begin{equation}\label{ka_comp}
        k_a^\mu = \omega_a u^\mu + \sigma_a |{\bf k}| \epsilon_L{}^{\mu}({\bf k}) \quad (\sigma_a = \pm 1) \,, 
    \end{equation}
and thus
    \begin{align}\label{eq:ppmm}
        \epsilon^{\mu\alpha\nu\beta}k_{1\alpha}k_{2\beta} 
        \propto \epsilon^{\mu\alpha\nu\beta}u_{\alpha}\epsilon_{L\beta} 
        \propto {\epsilon^\ast}_+{}^{\mu}\epsilon_+{}^{\nu} - {\epsilon^\ast}_-{}^{\mu}\epsilon_-{}^{\nu} \,. 
    \end{align}
The coefficient is a function of the cosmic time $t_c$. 
Therefore, it causes the isotropic difference in propagation between the left- and right-handed photons in the CMB rest frame. 

We can perform a similar argument to show that the operator ${\bf E}_\parallel \cdot {\bf B}_\parallel$ in Eq.~\eqref{op:aniso pv} generates the ACB. 
Recovering $u^\mu$ and $\bar{B}^\mu$, we can write it in a covariant way as
    \begin{align}\label{parallel}
        (\bar{B}_\mu u_\nu F^{\mu\nu} ) (\bar{B}_\alpha u_\beta \tilde{F}^{\alpha\beta} ) \,.
    \end{align}
Using the decomposition,
    \begin{align}\label{bB decomp}
    \bar{B}^\mu = \bar{B}_L \epsilon_L{}^\mu({\bf k}) + \bar{B}_+ \epsilon_+{}^\mu({\bf k}) + \bar{B}_- \epsilon_-{}^\mu({\bf k}) \,,
    \end{align}
we find
    \begin{align}
        \bar{B}_\mu u_\nu F^{\mu\nu} &\propto \left[ (\bar{B} \cdot k) u^\mu - (u \cdot k)\bar{B}^\mu \right] A_\mu \,, \\
        \bar{B}_\alpha u_\beta \tilde{F}^{\alpha\beta} &\propto \left[ \bar{B}_+ \epsilon_+{}^\alpha({\bf k}) - \bar{B}_- \epsilon_-{}^\alpha({\bf k}) \right] A_\alpha \,,
    \end{align}
and thus the operator of (\ref{parallel}) contains the term that causes the birefringence,
    \begin{align}\label{mag_cb}
         \bar{B}_+ \bar{B}_- \left({\epsilon^\ast}_+{}^{\mu}\epsilon_+{}^{\nu} - {\epsilon^\ast}_-{}^{\mu}\epsilon_-{}^{\nu} \right) A_{\mu}A_{\nu} \,.
    \end{align}
The amplitude $\bar{B}_+ \bar{B}_-$ represents the components of the magnetic field orthogonal to ${\bf k} \propto {\bf n}_{\rm LoS}$ with the line-of-sight direction ${\bf n}_{\rm LoS}$.
Therefore, the resultant birefringence angle depends on a particular direction.

\vspace{0.5cm}

\section{Dipole moment interactions}
\label{dipole op}

We here show that the operator (\ref{op:jf})
    \begin{align}
        J_{\mu\nu} F^{\mu\nu} \,, 
    \end{align}
induce the frequency-dependent ICB angle in the framework of the SMEFT/LEFT. 
In the SMEFT/LEFT, the only possibility is given by 
\cite{Giunti:2014ixa, Jenkins:2017jig, Altmannshofer:2018xyo}
    \begin{align}\label{op:dipole}
        J_{\mu\nu} = \bar{\nu}^i \sigma_{\mu\nu}\lambda^{ij} \nu^{j} \;;~ \lambda^{ij} \equiv \mu^{ij} + i \varepsilon^{ij} \gamma^5 \,,
    \end{align}
for three generations of neutrinos $\nu^i$ ($i=1,2,3$) with $\sigma_{\mu\nu} \equiv (i/2)[\gamma^\mu,\gamma^\nu]$, which corresponds to the magnetic ($\mu$) and electric ($\varepsilon$) dipole moment interactions of neutrinos. 
We assume that the neutrinos are Dirac fermions. For Majorana neutrinos, an extra factor of 1/2 should be included. 

To see how the operator (\ref{op:jf}) affects the propagation of a CMB photon, let us first write down the Maxwell equation: 
    \begin{align}
        \partial_\mu F^{\mu\nu} = -\partial_\mu J^{\mu\nu} \,.
    \end{align}
At the background level, the source term $\partial_\mu J^{\mu\nu}$ is independent of the electromagnetic field $A_\mu$ and thus does not modify its dispersion relation. 
Therefore, we need to consider the backreaction of $A_\mu$ to the neutrino field $\nu$. 

The Dirac equations of the neutrino fields in the mass eigenstates are modified as
    \begin{align}
        (i\slashed{\partial} - m_i)\nu^i = ( \sigma_{\mu\nu} \lambda^{ij} \nu^j ) F^{\mu\nu} \,. 
    \end{align}
These equations can be formally solved as
    \begin{align}\label{eq:nu formal sol}
        \nu^i = 
        (\nu^i)^{\rm (bg)} + (i\slashed{\partial}-m_i)^{-1} [( \sigma_{\mu\nu} \lambda^{ij} \nu^j ) F^{\mu\nu}] \,, 
    \end{align}
where $(i\slashed{\partial}-m_i)^{-1}$ is the inverse of $(i\slashed{\partial}-m_i)$. 
Here, $(\nu^i)^{\rm (bg)}$ is the homogeneous solution. 
It corresponds to the solution in the absence of $A_\mu$ and thus the background solution. 
Perturbatively expanding the solution (\ref{eq:nu formal sol}) in terms of $A_\mu$, 
we obtain
    \begin{align}\label{eq:nu sol}
        \nu^i = (\nu^i)^{\rm (bg)} + \hat{\psi}_{\mu\nu}^i F^{\mu\nu} + {\cal O}(A_\mu^2) \,,
    \end{align}
where 
    \begin{align}
        \hat{\psi}_{\mu\nu}^i \equiv (i\slashed{\partial}-m_i)^{-1}  \sigma_{\mu\nu} \lambda^{ij} (\nu^j)^{\rm (bg)}  \,.
    \end{align}
Hereafter, a quantity with a hat denotes an operator on the electromagnetic field $F^{\mu\nu}$. 
Substituting this solution into (\ref{op:dipole}), we find
    \begin{align}\label{eq:jkf}
     \delta J_{\mu\nu} = \hat{K}_{\mu\nu\alpha\beta}F^{\alpha\beta} \,,
    \end{align}
where
     \begin{align}
        \hat{K}_{\mu\nu\alpha\beta}
        \equiv (\bar{\nu}^i)^{\rm (bg)} (\sigma_{\mu\nu} \lambda^{ij}) \hat{\psi}_{\alpha\beta}^j + {\rm h.c.} \,.
    \end{align}
In terms of $\hat{K}_{\mu\nu\alpha\beta}$, the source term in the Maxwell equation is written as
     \begin{align}
        -\partial_\mu (\delta J^{\mu\nu}) = -\partial_\mu (\hat{K}^{\mu\nu}{}_{\alpha\beta}F^{\alpha\beta})  \,,
    \end{align}
and can be derived from the following interaction in the Lagrangian density 
    \begin{align}\label{eq:fkf}
        {\cal L}_{\rm eff} = 
        -\frac{1}{4}
        F^{\mu\nu}\hat{K}_{\mu\nu\alpha\beta}F^{\alpha\beta} \,.
    \end{align}
In the action, $\hat{K}_{\mu\nu\alpha\beta}$ can be replaced as
     \begin{align}
        &\hat{K}_{\mu\nu\alpha\beta} \to
        2(\bar{\nu}^i)^{\rm (bg)} (\sigma_{\mu\nu} \lambda^{ij}) (i\slashed{\partial}-m_j)^{-1} (\sigma_{\alpha\beta}\lambda^{jk})(\nu^k)^{\rm (bg)} \,.
    \end{align}
Hereafter, we will use this expression for $\hat{K}_{\mu\nu\alpha\beta}$. 

To pick up the CS-type operator, 
we would like to extract the following components,
    \begin{align}\label{eq:oep}
         \hat{K}_{\mu\nu\alpha\beta} \supset \frac{\hat{\cal O}_{\epsilon} \epsilon_{\mu\nu\alpha\beta}}{2} \,,
    \end{align}
which results in the CS-type operator
    \begin{align}\label{op:OeCS app}
        -\frac{1}{4}F_{\mu\nu} \hat{\cal O}_{\epsilon} \tilde{F}^{\mu\nu} \,.
    \end{align}
From the background symmetry, 
the interaction should be written as
    \begin{align}
         \int {\rm d}^3 x ~ F_{\mu\nu} \hat{\cal O}_{\epsilon} \tilde{F}^{\mu\nu} = \int \frac{{\rm d}^3 p_\gamma}{(2\pi)^3} \tilde{\cal O}_{\epsilon}(\omega) F_{\mu\nu} (-p_\gamma) \tilde{F}^{\mu\nu}(p_\gamma) \,,
    \end{align}
where $\omega$ is the frequency of the CMB photon: $p_\gamma = (\omega, \omega {\bf n})$. 
Here, we have assumed that the cosmic expansion is adiabatic, which is appropriate for the CMB photons. 
In the following, we will derive the expression of $\tilde{\cal O}_{\epsilon}(\omega)$. 

We can extract the component (\ref{eq:oep}) by contracting $\hat{K}_{\mu\nu\alpha\beta}$ with $\epsilon^{\mu\nu\alpha\beta}$: 
    \begin{align}
        \hat{\cal O}_{\epsilon} = -\frac{\epsilon^{\mu\nu\alpha\beta}\hat{K}_{\mu\nu\alpha\beta}}{12} \,.
    \end{align}
Using the identities $\epsilon^{\mu\nu\alpha\beta} \sigma_{\alpha\beta} = -2i\gamma^5 \sigma^{\mu\nu}$, 
$\sigma_{\mu\nu} \sigma^{\mu\nu} = 12$ and $\sigma_{\mu\nu} \gamma^\alpha \sigma^{\mu\nu} = 0$,
$\hat{\cal O}_{\epsilon}$ can be computed as
    \begin{align}
        \hat{\cal O}_{\epsilon} 
        = -4im_j (\bar{\nu}^i)^{\rm (bg)} \lambda^{ij} \gamma^5 (
        \partial^2 + m_j^2
        )^{-1} \lambda^{jk} (\nu^k)^{\rm (bg)} \,.
    \end{align}
Replacing the background neutrino bilinears in the momentum space ($p_\nu = (E_\nu, {\bf p}_\nu)$) with the expectation values as $(\bar{\nu}^i)^{\rm (bg)}(p'_\nu) \gamma^5 (\nu^k)^{\rm (bg)}(p_\nu) \to 0$ and 
    \begin{align}
        &(\bar{\nu}^i)^{\rm (bg)}(p'_\nu) (\nu^k)^{\rm (bg)}(p_\nu) \to
        \delta^{ik}(m_i/E_\nu)[n_\nu^i(p_\nu) + \bar{n}_\nu^i(p_\nu)](2\pi)^3\delta^{(3)}({\bf p}'_\nu - {\bf p}_\nu)
    \end{align}
(see appendix \ref{backgroundnu}), 
we can read
    \begin{align}
        &\tilde{\cal O}_{\epsilon}(\omega)
        = -2(\mu^{ij} \varepsilon^{ji} + \varepsilon^{ij}\mu^{ji} ) m_j 
        \int \frac{{\rm d}^3 p_\nu}{(2\pi)^3} \frac{m_i}{E_\nu}\frac{[n_\nu^i(p_\nu) + \bar{n}_\nu^i(p_\nu)]}{(p_\nu + p_\gamma)^2-m_j^2} \,.
    \end{align}
Taking into account that $p_\nu$ is the momentum of the $i$-th neutrino, we can explicitly write the denominator in the integrand as,
    \begin{align}\label{eq:denominator}
        (p_\nu + p_\gamma)^2-m_j^2 = 2\omega(E_\nu - {\bf n} \cdot {\bf p}_\nu) + m_i^2 - m_j^2 \,.
    \end{align}
Since the neutrino number density quickly decays due to the cosmic expansion, we can well approximate the ICB angle by $-\tilde{\cal O}_{\epsilon}(\omega)/2$ at the time of last scattering. 
Thus, $\omega$ is estimated as $\omega \sim T_{\rm LSS} \sim 0.3 {\rm eV}$ and the first term is dominant in (\ref{eq:denominator}):
    \begin{align}
        (p_\nu + p_\gamma)^2-m_j^2 \simeq 2\omega(E_\nu - {\bf n} \cdot {\bf p}_\nu) \,.
    \end{align}
In conclusion, we find
    \begin{align}
        \beta \simeq -\left. \frac{\tilde{\cal O}_{\epsilon}(\omega)}{2} \right|_{t = t_{\rm LSS}}
        &\propto \frac{1}{\omega} \,,
    \end{align}
and thus the operator (\ref{op:OeCS app}) induces the frequency-dependent ICB angle. 

\vspace{0.5cm}

\section{Background neutrinos}
\label{backgroundnu}

We here consider the cosmic background neutrino $\nu$ as a free Dirac fermion with mass $m$,
satisfying the Dirac equation, $(i\slashed{\partial} - m)\nu=0$.
The cosmic expansion is adiabatic, and hence the (gravitational) particle production is a sub-leading effect. 
Quantization of the Dirac field gives 
    \begin{align}
        \nu(x)&=\int \frac{{d}^3p}{(2\pi)^3} \frac{1}{\sqrt{2E_{\mathbf{p}}}} \sum_s\left[a_\mathbf{p}^s u^s(p) e^{-ipx}+b_{\mathbf{p}}^{s\dagger} v^s(p) e^{ipx}\right] , \\
        \bar{\nu}(x)&=\int \frac{{d}^3p}{(2\pi)^3} \frac{1}{\sqrt{2E_{\mathbf{p}}}} \sum_s\left[b_{\mathbf{p}}^s \bar{v}^s(p) e^{-ipx}+a_{\mathbf{p}}^{s \dagger}\bar{u}^s(p)e^{ipx}\right] .
    \end{align}
Here, $a_\mathbf{p}^s, b_{\mathbf{p}}^s$ $(s=1,2)$ denote the operator coefficients.
They satisfy the anti-commutation relations,
$\{a_{\mathbf{p}}^r , a_{\mathbf{q}}^{s\dagger}\} = \{ b_{\mathbf{p}}^r , b_{\mathbf{q}}^{s\dagger} \}
= (2\pi)^3 \delta^{(3)} (\mathbf{p}-\mathbf{q}) \delta^{rs}$
and all the other anti-commutators are equal to zero.
Spinor functions $u, v$ are given by solutions of the Dirac equation,
    \begin{equation}
        u^s(p)=
        \begin{pmatrix}
            \sqrt{p\cdot\beta}\, \xi^s\\
            \sqrt{p\cdot\bar{\beta}}\, \xi^s
        \end{pmatrix} , \quad 
        v^s(p)=
        \begin{pmatrix}
            \sqrt{p\cdot\beta}\, \eta^s\\
            -\sqrt{p\cdot\bar{\beta}}\, \eta^s
        \end{pmatrix} ,
    \end{equation}
where $\beta^\mu \equiv (1, \boldsymbol{\beta})$ and $\bar{\beta}^\mu \equiv (1, -\boldsymbol{\beta})$
with Pauli matrices $\boldsymbol{\beta}$.
The vectors $\xi$ and $\eta$ are both two-component spinors normalized as $\xi^\dagger\xi=\eta^\dagger\eta=1$.
Then, the expectation value of $\bar{\nu} \nu$ with regard to a state of fixed neutrino and anti-neutrino number densities
are obtained as
    \begin{align}
        \langle \bar{\nu} \nu \rangle
        = \int \frac{{d}^3 p}{\left( 2 \pi \right)^3} \frac{m}{E_{\mathbf{p}}}\left[n(p,t)+\bar{n}(p,t)\right] ,
    \end{align}
where $n, \bar{n}$ denote the number densities of the neutrino and anti-neutrino, respectively,
and we have used $\bar{u}^s u^r= 2m\delta^{sr}$ and $\bar{v}^s v^r=-2m\delta^{sr}$.
On the other hand,
using $\bar{u}^s\gamma^5 u^r = 0$ and $\bar{v}^s \gamma^5 v^r =0$,
we find $\langle \bar{\nu} \gamma^5 \nu \rangle = 0$.
For a Majorana neutrino, one can simply take $a_{\mathbf{p}}=b_{\mathbf{p}}$ and
the similar expectation values are derived.


\providecommand{\href}[2]{#2}\begingroup\raggedright\endgroup



\begin{thebibliography}{10}

\bibitem{WMAP:2003elm}
{\bfseries WMAP} Collaboration, D.~N. Spergel {\em et~al.}, ``{First year
  Wilkinson Microwave Anisotropy Probe (WMAP) observations: Determination of
  cosmological parameters},'' \href{http://dx.doi.org/10.1086/377226}{{\em
  Astrophys. J. Suppl.} {\bfseries 148} (2003) 175--194},
  \href{http://arxiv.org/abs/astro-ph/0302209}{{\ttfamily
  arXiv:astro-ph/0302209}}.

\bibitem{Komatsu:2014ioa}
{\bfseries WMAP Science Team} Collaboration, E.~Komatsu and C.~L. Bennett,
  ``{Results from the Wilkinson Microwave Anisotropy Probe},''
  \href{http://dx.doi.org/10.1093/ptep/ptu083}{{\em PTEP} {\bfseries 2014}
  (2014) 06B102}, \href{http://arxiv.org/abs/1404.5415}{{\ttfamily
  arXiv:1404.5415 [astro-ph.CO]}}.

\bibitem{Planck:2013pxb}
{\bfseries Planck} Collaboration, P.~A.~R. Ade {\em et~al.}, ``{Planck 2013
  results. XVI. Cosmological parameters},''
  \href{http://dx.doi.org/10.1051/0004-6361/201321591}{{\em Astron. Astrophys.}
  {\bfseries 571} (2014) A16}, \href{http://arxiv.org/abs/1303.5076}{{\ttfamily
  arXiv:1303.5076 [astro-ph.CO]}}.

\bibitem{Planck:2018vyg}
{\bfseries Planck} Collaboration, N.~Aghanim {\em et~al.}, ``{Planck 2018
  results. VI. Cosmological parameters},''
  \href{http://dx.doi.org/10.1051/0004-6361/201833910}{{\em Astron. Astrophys.}
  {\bfseries 641} (2020) A6}, \href{http://arxiv.org/abs/1807.06209}{{\ttfamily
  arXiv:1807.06209 [astro-ph.CO]}}. [Erratum: Astron.Astrophys. 652, C4
  (2021)].

\bibitem{Minami:2020odp}
Y.~Minami and E.~Komatsu, ``{New Extraction of the Cosmic Birefringence from
  the Planck 2018 Polarization Data},''
  \href{http://dx.doi.org/10.1103/PhysRevLett.125.221301}{{\em Phys. Rev.
  Lett.} {\bfseries 125} no.~22, (2020) 221301},
  \href{http://arxiv.org/abs/2011.11254}{{\ttfamily arXiv:2011.11254
  [astro-ph.CO]}}.

\bibitem{Diego-Palazuelos:2022dsq}
P.~Diego-Palazuelos {\em et~al.}, ``{Cosmic Birefringence from the Planck Data
  Release 4},'' \href{http://dx.doi.org/10.1103/PhysRevLett.128.091302}{{\em
  Phys. Rev. Lett.} {\bfseries 128} no.~9, (2022) 091302},
  \href{http://arxiv.org/abs/2201.07682}{{\ttfamily arXiv:2201.07682
  [astro-ph.CO]}}.

\bibitem{Eskilt:2022wav}
J.~R. Eskilt, ``{Frequency-dependent constraints on cosmic birefringence from
  the LFI and HFI Planck Data Release 4},''
  \href{http://dx.doi.org/10.1051/0004-6361/202243269}{{\em Astron. Astrophys.}
  {\bfseries 662} (2022) A10},
  \href{http://arxiv.org/abs/2201.13347}{{\ttfamily arXiv:2201.13347
  [astro-ph.CO]}}.

\bibitem{Eskilt:2022cff}
J.~R. Eskilt and E.~Komatsu, ``{Improved Constraints on Cosmic Birefringence
  from the WMAP and Planck Cosmic Microwave Background Polarization Data},''
  \href{http://arxiv.org/abs/2205.13962}{{\ttfamily arXiv:2205.13962
  [astro-ph.CO]}}.

\bibitem{Eskilt:2023ndm}
J.~R. Eskilt {\em et~al.}, ``{Cosmoglobe DR1 results. II. Constraints on
  isotropic cosmic birefringence from reprocessed WMAP and Planck LFI data},''
  \href{http://arxiv.org/abs/2305.02268}{{\ttfamily arXiv:2305.02268
  [astro-ph.CO]}}.

\bibitem{Carroll:1989vb}
S.~M. Carroll, G.~B. Field, and R.~Jackiw, ``{Limits on a Lorentz and Parity
  Violating Modification of Electrodynamics},''
  \href{http://dx.doi.org/10.1103/PhysRevD.41.1231}{{\em Phys. Rev. D}
  {\bfseries 41} (1990) 1231}.

\bibitem{Carroll:1991zs}
S.~M. Carroll and G.~B. Field, ``{The Einstein equivalence principle and the
  polarization of radio galaxies},''
  \href{http://dx.doi.org/10.1103/PhysRevD.43.3789}{{\em Phys. Rev. D}
  {\bfseries 43} (1991) 3789}.

\bibitem{Harari:1992ea}
D.~Harari and P.~Sikivie, ``{Effects of a Nambu-Goldstone boson on the
  polarization of radio galaxies and the cosmic microwave background},''
  \href{http://dx.doi.org/10.1016/0370-2693(92)91363-E}{{\em Phys. Lett. B}
  {\bfseries 289} (1992) 67--72}.

\bibitem{Feng:2006dp}
B.~Feng, M.~Li, J.-Q. Xia, X.~Chen, and X.~Zhang, ``{Searching for CPT
  Violation with Cosmic Microwave Background Data from WMAP and BOOMERANG},''
  \href{http://dx.doi.org/10.1103/PhysRevLett.96.221302}{{\em Phys. Rev. Lett.}
  {\bfseries 96} (2006) 221302},
  \href{http://arxiv.org/abs/astro-ph/0601095}{{\ttfamily
  arXiv:astro-ph/0601095}}.

\bibitem{QUaD:2008ado}
{\bfseries QUaD} Collaboration, E.~Y.~S. Wu {\em et~al.}, ``{Parity Violation
  Constraints Using Cosmic Microwave Background Polarization Spectra from 2006
  and 2007 Observations by the QUaD Polarimeter},''
  \href{http://dx.doi.org/10.1103/PhysRevLett.102.161302}{{\em Phys. Rev.
  Lett.} {\bfseries 102} (2009) 161302},
  \href{http://arxiv.org/abs/0811.0618}{{\ttfamily arXiv:0811.0618
  [astro-ph]}}.

\bibitem{WMAP:2010qai}
{\bfseries WMAP} Collaboration, E.~Komatsu {\em et~al.}, ``{Seven-Year
  Wilkinson Microwave Anisotropy Probe (WMAP) Observations: Cosmological
  Interpretation},'' \href{http://dx.doi.org/10.1088/0067-0049/192/2/18}{{\em
  Astrophys. J. Suppl.} {\bfseries 192} (2011) 18},
  \href{http://arxiv.org/abs/1001.4538}{{\ttfamily arXiv:1001.4538
  [astro-ph.CO]}}.

\bibitem{Planck:2016soo}
{\bfseries Planck} Collaboration, N.~Aghanim {\em et~al.}, ``{Planck
  intermediate results. XLIX. Parity-violation constraints from polarization
  data},'' \href{http://dx.doi.org/10.1051/0004-6361/201629018}{{\em Astron.
  Astrophys.} {\bfseries 596} (2016) A110},
  \href{http://arxiv.org/abs/1605.08633}{{\ttfamily arXiv:1605.08633
  [astro-ph.CO]}}.

\bibitem{Minami:2019ruj}
Y.~Minami, H.~Ochi, K.~Ichiki, N.~Katayama, E.~Komatsu, and T.~Matsumura,
  ``{Simultaneous determination of the cosmic birefringence and miscalibrated
  polarization angles from CMB experiments},''
  \href{http://dx.doi.org/10.1093/ptep/ptz079}{{\em PTEP} {\bfseries 2019}
  no.~8, (2019) 083E02}, \href{http://arxiv.org/abs/1904.12440}{{\ttfamily
  arXiv:1904.12440 [astro-ph.CO]}}.

\bibitem{Minami:2020xfg}
Y.~Minami, ``{Determination of miscalibrated polarization angles from observed
  cosmic microwave background and foreground $EB$ power spectra: Application to
  partial-sky observation},''
  \href{http://dx.doi.org/10.1093/ptep/ptaa057}{{\em PTEP} {\bfseries 2020}
  no.~6, (2020) 063E01}, \href{http://arxiv.org/abs/2002.03572}{{\ttfamily
  arXiv:2002.03572 [astro-ph.CO]}}.

\bibitem{Minami:2020fin}
Y.~Minami and E.~Komatsu, ``{Simultaneous determination of the cosmic
  birefringence and miscalibrated polarization angles II: Including cross
  frequency spectra},'' \href{http://dx.doi.org/10.1093/ptep/ptaa130}{{\em
  PTEP} {\bfseries 2020} no.~10, (2020) 103E02},
  \href{http://arxiv.org/abs/2006.15982}{{\ttfamily arXiv:2006.15982
  [astro-ph.CO]}}.

\bibitem{Clark:2021kze}
S.~E. Clark, C.-G. Kim, J.~C. Hill, and B.~S. Hensley, ``{The Origin of Parity
  Violation in Polarized Dust Emission and Implications for Cosmic
  Birefringence},'' \href{http://dx.doi.org/10.3847/1538-4357/ac0e35}{{\em
  Astrophys. J.} {\bfseries 919} no.~1, (2021) 53},
  \href{http://arxiv.org/abs/2105.00120}{{\ttfamily arXiv:2105.00120
  [astro-ph.GA]}}.

\bibitem{Diego-Palazuelos:2022cnh}
P.~Diego-Palazuelos {\em et~al.}, ``{Robustness of cosmic birefringence
  measurement against Galactic foreground emission and instrumental
  systematics},'' \href{http://dx.doi.org/10.1088/1475-7516/2023/01/044}{{\em
  JCAP} {\bfseries 01} (2023) 044},
  \href{http://arxiv.org/abs/2210.07655}{{\ttfamily arXiv:2210.07655
  [astro-ph.CO]}}.

\bibitem{Monelli:2022pru}
M.~Monelli, E.~Komatsu, A.~E. Adler, M.~Billi, P.~Campeti, N.~Dachlythra, A.~J.
  Duivenvoorden, J.~E. Gudmundsson, and M.~Reinecke, ``{Impact of half-wave
  plate systematics on the measurement of cosmic birefringence from CMB
  polarization},'' \href{http://arxiv.org/abs/2211.05685}{{\ttfamily
  arXiv:2211.05685 [astro-ph.CO]}}.

\bibitem{Jost:2022oab}
B.~Jost, J.~Errard, and R.~Stompor, ``{Characterising cosmic birefringence in
  the presence of galactic foregrounds and instrumental systematic effects},''
  \href{http://arxiv.org/abs/2212.08007}{{\ttfamily arXiv:2212.08007
  [astro-ph.CO]}}.

\bibitem{Komatsu:2022nvu}
E.~Komatsu, ``{New physics from the polarized light of the cosmic microwave
  background},'' \href{http://dx.doi.org/10.1038/s42254-022-00452-4}{{\em
  Nature Rev. Phys.} {\bfseries 4} no.~7, (2022) 452--469},
  \href{http://arxiv.org/abs/2202.13919}{{\ttfamily arXiv:2202.13919
  [astro-ph.CO]}}.

\bibitem{Carroll:1998zi}
S.~M. Carroll, ``{Quintessence and the rest of the world},''
  \href{http://dx.doi.org/10.1103/PhysRevLett.81.3067}{{\em Phys. Rev. Lett.}
  {\bfseries 81} (1998) 3067--3070},
  \href{http://arxiv.org/abs/astro-ph/9806099}{{\ttfamily
  arXiv:astro-ph/9806099}}.

\bibitem{Lue:1998mq}
A.~Lue, L.-M. Wang, and M.~Kamionkowski, ``{Cosmological signature of new
  parity violating interactions},''
  \href{http://dx.doi.org/10.1103/PhysRevLett.83.1506}{{\em Phys. Rev. Lett.}
  {\bfseries 83} (1999) 1506--1509},
  \href{http://arxiv.org/abs/astro-ph/9812088}{{\ttfamily
  arXiv:astro-ph/9812088}}.

\bibitem{Pospelov:2008gg}
M.~Pospelov, A.~Ritz, C.~Skordis, A.~Ritz, and C.~Skordis, ``{Pseudoscalar
  perturbations and polarization of the cosmic microwave background},''
  \href{http://dx.doi.org/10.1103/PhysRevLett.103.051302}{{\em Phys. Rev.
  Lett.} {\bfseries 103} (2009) 051302},
  \href{http://arxiv.org/abs/0808.0673}{{\ttfamily arXiv:0808.0673
  [astro-ph]}}.

\bibitem{Finelli:2008jv}
F.~Finelli and M.~Galaverni, ``{Rotation of Linear Polarization Plane and
  Circular Polarization from Cosmological Pseudo-Scalar Fields},''
  \href{http://dx.doi.org/10.1103/PhysRevD.79.063002}{{\em Phys. Rev. D}
  {\bfseries 79} (2009) 063002},
  \href{http://arxiv.org/abs/0802.4210}{{\ttfamily arXiv:0802.4210
  [astro-ph]}}.

\bibitem{Panda:2010uq}
S.~Panda, Y.~Sumitomo, and S.~P. Trivedi, ``{Axions as Quintessence in String
  Theory},'' \href{http://dx.doi.org/10.1103/PhysRevD.83.083506}{{\em Phys.
  Rev. D} {\bfseries 83} (2011) 083506},
  \href{http://arxiv.org/abs/1011.5877}{{\ttfamily arXiv:1011.5877 [hep-th]}}.

\bibitem{Lee:2013mqa}
S.~Lee, G.-C. Liu, and K.-W. Ng, ``{Imprint of Scalar Dark Energy on Cosmic
  Microwave Background Polarization},''
  \href{http://dx.doi.org/10.1103/PhysRevD.89.063010}{{\em Phys. Rev. D}
  {\bfseries 89} no.~6, (2014) 063010},
  \href{http://arxiv.org/abs/1307.6298}{{\ttfamily arXiv:1307.6298
  [astro-ph.CO]}}.

\bibitem{Zhao:2014yna}
W.~Zhao and M.~Li, ``{Fluctuations of cosmological birefringence and the effect
  on CMB B-mode polarization},''
  \href{http://dx.doi.org/10.1103/PhysRevD.89.103518}{{\em Phys. Rev. D}
  {\bfseries 89} no.~10, (2014) 103518},
  \href{http://arxiv.org/abs/1403.3997}{{\ttfamily arXiv:1403.3997
  [astro-ph.CO]}}.

\bibitem{Liu:2016dcg}
G.-C. Liu and K.-W. Ng, ``{Axion Dark Matter Induced Cosmic Microwave
  Background $B$-modes},''
  \href{http://dx.doi.org/10.1016/j.dark.2017.02.004}{{\em Phys. Dark Univ.}
  {\bfseries 16} (2017) 22--25},
  \href{http://arxiv.org/abs/1612.02104}{{\ttfamily arXiv:1612.02104
  [astro-ph.CO]}}.

\bibitem{Sigl:2018fba}
G.~Sigl and P.~Trivedi, ``{Axion-like Dark Matter Constraints from CMB
  Birefringence},'' \href{http://arxiv.org/abs/1811.07873}{{\ttfamily
  arXiv:1811.07873 [astro-ph.CO]}}.

\bibitem{Fedderke:2019ajk}
M.~A. Fedderke, P.~W. Graham, and S.~Rajendran, ``{Axion Dark Matter Detection
  with CMB Polarization},''
  \href{http://dx.doi.org/10.1103/PhysRevD.100.015040}{{\em Phys. Rev. D}
  {\bfseries 100} no.~1, (2019) 015040},
  \href{http://arxiv.org/abs/1903.02666}{{\ttfamily arXiv:1903.02666
  [astro-ph.CO]}}.

\bibitem{Fujita:2020ecn}
T.~Fujita, K.~Murai, H.~Nakatsuka, and S.~Tsujikawa, ``{Detection of isotropic
  cosmic birefringence and its implications for axionlike particles including
  dark energy},'' \href{http://dx.doi.org/10.1103/PhysRevD.103.043509}{{\em
  Phys. Rev. D} {\bfseries 103} no.~4, (2021) 043509},
  \href{http://arxiv.org/abs/2011.11894}{{\ttfamily arXiv:2011.11894
  [astro-ph.CO]}}.

\bibitem{Takahashi:2020tqv}
F.~Takahashi and W.~Yin, ``{Kilobyte Cosmic Birefringence from ALP Domain
  Walls},'' \href{http://dx.doi.org/10.1088/1475-7516/2021/04/007}{{\em JCAP}
  {\bfseries 04} (2021) 007}, \href{http://arxiv.org/abs/2012.11576}{{\ttfamily
  arXiv:2012.11576 [hep-ph]}}.

\bibitem{Fung:2021wbz}
L.~W.~H. Fung, L.~Li, T.~Liu, H.~N. Luu, Y.-C. Qiu, and S.~H.~H. Tye,
  ``{Axi-Higgs cosmology},''
  \href{http://dx.doi.org/10.1088/1475-7516/2021/08/057}{{\em JCAP} {\bfseries
  08} (2021) 057}, \href{http://arxiv.org/abs/2102.11257}{{\ttfamily
  arXiv:2102.11257 [hep-ph]}}.

\bibitem{Nakagawa:2021nme}
S.~Nakagawa, F.~Takahashi, and M.~Yamada, ``{Cosmic Birefringence Triggered by
  Dark Matter Domination},''
  \href{http://dx.doi.org/10.1103/PhysRevLett.127.181103}{{\em Phys. Rev.
  Lett.} {\bfseries 127} no.~18, (2021) 181103},
  \href{http://arxiv.org/abs/2103.08153}{{\ttfamily arXiv:2103.08153
  [hep-ph]}}.

\bibitem{Jain:2021shf}
M.~Jain, A.~J. Long, and M.~A. Amin, ``{CMB birefringence from ultralight-axion
  string networks},''
  \href{http://dx.doi.org/10.1088/1475-7516/2021/05/055}{{\em JCAP} {\bfseries
  05} (2021) 055}, \href{http://arxiv.org/abs/2103.10962}{{\ttfamily
  arXiv:2103.10962 [astro-ph.CO]}}.

\bibitem{Choi:2021aze}
G.~Choi, W.~Lin, L.~Visinelli, and T.~T. Yanagida, ``{Cosmic birefringence and
  electroweak axion dark energy},''
  \href{http://dx.doi.org/10.1103/PhysRevD.104.L101302}{{\em Phys. Rev. D}
  {\bfseries 104} no.~10, (2021) L101302},
  \href{http://arxiv.org/abs/2106.12602}{{\ttfamily arXiv:2106.12602
  [hep-ph]}}.

\bibitem{Obata:2021nql}
I.~Obata, ``{Implications of the cosmic birefringence measurement for the axion
  dark matter search},''
  \href{http://dx.doi.org/10.1088/1475-7516/2022/09/062}{{\em JCAP} {\bfseries
  09} (2022) 062}, \href{http://arxiv.org/abs/2108.02150}{{\ttfamily
  arXiv:2108.02150 [astro-ph.CO]}}.

\bibitem{Nakatsuka:2022epj}
H.~Nakatsuka, T.~Namikawa, and E.~Komatsu, ``{Is cosmic birefringence due to
  dark energy or dark matter? A tomographic approach},''
  \href{http://dx.doi.org/10.1103/PhysRevD.105.123509}{{\em Phys. Rev. D}
  {\bfseries 105} no.~12, (2022) 123509},
  \href{http://arxiv.org/abs/2203.08560}{{\ttfamily arXiv:2203.08560
  [astro-ph.CO]}}.

\bibitem{Lin:2022niw}
W.~Lin and T.~T. Yanagida, ``{Consistency of the string inspired electroweak
  axion with cosmic birefringence},''
  \href{http://dx.doi.org/10.1103/PhysRevD.107.L021302}{{\em Phys. Rev. D}
  {\bfseries 107} no.~2, (2023) L021302},
  \href{http://arxiv.org/abs/2208.06843}{{\ttfamily arXiv:2208.06843
  [hep-ph]}}.

\bibitem{Gasparotto:2022uqo}
S.~Gasparotto and I.~Obata, ``{Cosmic birefringence from monodromic axion dark
  energy},'' \href{http://dx.doi.org/10.1088/1475-7516/2022/08/025}{{\em JCAP}
  {\bfseries 08} no.~08, (2022) 025},
  \href{http://arxiv.org/abs/2203.09409}{{\ttfamily arXiv:2203.09409
  [astro-ph.CO]}}.

\bibitem{Lee:2022udm}
N.~Lee, S.~C. Hotinli, and M.~Kamionkowski, ``{Probing cosmic birefringence
  with polarized Sunyaev-Zel\textquoteright{}dovich tomography},''
  \href{http://dx.doi.org/10.1103/PhysRevD.106.083518}{{\em Phys. Rev. D}
  {\bfseries 106} no.~8, (2022) 083518},
  \href{http://arxiv.org/abs/2207.05687}{{\ttfamily arXiv:2207.05687
  [astro-ph.CO]}}.

\bibitem{Jain:2022jrp}
M.~Jain, R.~Hagimoto, A.~J. Long, and M.~A. Amin, ``{Searching for axion-like
  particles through CMB birefringence from string-wall networks},''
  \href{http://dx.doi.org/10.1088/1475-7516/2022/10/090}{{\em JCAP} {\bfseries
  10} (2022) 090}, \href{http://arxiv.org/abs/2208.08391}{{\ttfamily
  arXiv:2208.08391 [astro-ph.CO]}}.

\bibitem{Murai:2022zur}
K.~Murai, F.~Naokawa, T.~Namikawa, and E.~Komatsu, ``{Isotropic cosmic
  birefringence from early dark energy},''
  \href{http://dx.doi.org/10.1103/PhysRevD.107.L041302}{{\em Phys. Rev. D}
  {\bfseries 107} no.~4, (2023) L041302},
  \href{http://arxiv.org/abs/2209.07804}{{\ttfamily arXiv:2209.07804
  [astro-ph.CO]}}.

\bibitem{Gonzalez:2022mcx}
D.~Gonzalez, N.~Kitajima, F.~Takahashi, and W.~Yin, ``{Stability of domain wall
  network with initial inflationary fluctuations and its implications for
  cosmic birefringence},''
  \href{http://dx.doi.org/10.1016/j.physletb.2023.137990}{{\em Phys. Lett. B}
  {\bfseries 843} (2023) 137990},
  \href{http://arxiv.org/abs/2211.06849}{{\ttfamily arXiv:2211.06849
  [hep-ph]}}.

\bibitem{Qiu:2023los}
Y.-C. Qiu, J.-W. Wang, and T.~T. Yanagida, ``{High-quality axions in a class of
  chiral $U(1)$ gauge theories},''
  \href{http://arxiv.org/abs/2301.02345}{{\ttfamily arXiv:2301.02345
  [hep-ph]}}.

\bibitem{Eskilt:2023nxm}
J.~R. Eskilt, L.~Herold, E.~Komatsu, K.~Murai, T.~Namikawa, and F.~Naokawa,
  ``{Constraints on Early Dark Energy from Isotropic Cosmic Birefringence},''
  \href{http://dx.doi.org/10.1103/PhysRevLett.131.121001}{{\em Phys. Rev.
  Lett.} {\bfseries 131} no.~12, (2023) 121001},
  \href{http://arxiv.org/abs/2303.15369}{{\ttfamily arXiv:2303.15369
  [astro-ph.CO]}}.

\bibitem{Namikawa:2023zux}
T.~Namikawa and I.~Obata, ``{Cosmic birefringence tomography with polarized
  Sunyaev Zel'dovich effect},''
  \href{http://arxiv.org/abs/2306.08875}{{\ttfamily arXiv:2306.08875
  [astro-ph.CO]}}.

\bibitem{Gasparotto:2023psh}
S.~Gasparotto and E.~I. Sfakianakis, ``{Cosmic Birefringence from the
  Axiverse},'' \href{http://arxiv.org/abs/2306.16355}{{\ttfamily
  arXiv:2306.16355 [astro-ph.CO]}}.

\bibitem{Agrawal:2022lsp}
P.~Agrawal, M.~Nee, and M.~Reig, ``{Axion couplings in grand unified
  theories},'' \href{http://dx.doi.org/10.1007/JHEP10(2022)141}{{\em JHEP}
  {\bfseries 10} (2022) 141}, \href{http://arxiv.org/abs/2206.07053}{{\ttfamily
  arXiv:2206.07053 [hep-ph]}}.

\bibitem{Bartolo:2019eac}
N.~Bartolo, A.~Hoseinpour, S.~Matarrese, G.~Orlando, and M.~Zarei, ``{CMB
  Circular and B-mode Polarization from New Interactions},''
  \href{http://dx.doi.org/10.1103/PhysRevD.100.043516}{{\em Phys. Rev. D}
  {\bfseries 100} no.~4, (2019) 043516},
  \href{http://arxiv.org/abs/1903.04578}{{\ttfamily arXiv:1903.04578
  [hep-ph]}}.

\bibitem{Royer:1968rg}
J.~Royer, ``{EFFECT OF A DEGENERATE NEUTRINO SEA ON ELECTROMAGNETISM},''
  \href{http://dx.doi.org/10.1103/PhysRev.174.1719}{{\em Phys. Rev.} {\bfseries
  174} (1968) 1719--1724}.

\bibitem{Karl:1975df}
G.~Karl, ``{Coherent Parity Violation: A Review of Optical Activity with
  Massless and Massive Particles},''
  \href{http://dx.doi.org/10.1139/p76-061}{{\em Can. J. Phys.} {\bfseries 54}
  (1976) 568}.

\bibitem{Mohanty:1997mr}
S.~Mohanty, J.~F. Nieves, and P.~B. Pal, ``{Optical activity of a neutrino
  gas},'' \href{http://dx.doi.org/10.1103/PhysRevD.58.093007}{{\em Phys. Rev.
  D} {\bfseries 58} (1998) 093007},
  \href{http://arxiv.org/abs/hep-ph/9712414}{{\ttfamily arXiv:hep-ph/9712414}}.

\bibitem{Karl:2004bt}
G.~Karl and V.~Novikov, ``{Photon-neutrino interactions},''
  \href{http://dx.doi.org/10.1134/1.1931009}{{\em JETP Lett.} {\bfseries 81}
  (2005) 249--254}, \href{http://arxiv.org/abs/hep-ph/0411176}{{\ttfamily
  arXiv:hep-ph/0411176}}.

\bibitem{Dvornikov:2020olb}
M.~Dvornikov and V.~B. Semikoz, ``{Birefringence of electromagnetic waves in
  the relic neutrino gas},''
  \href{http://dx.doi.org/10.1088/1475-7516/2021/03/028}{{\em JCAP} {\bfseries
  03} (2021) 028}, \href{http://arxiv.org/abs/2011.14883}{{\ttfamily
  arXiv:2011.14883 [hep-ph]}}.

\bibitem{Petropavlova:2022spq}
M.~Petropavlova and A.~Smetana, ``{Toward interferometry of neutrino
  electromagnetism},''
  \href{http://dx.doi.org/10.1103/PhysRevD.106.053003}{{\em Phys. Rev. D}
  {\bfseries 106} no.~5, (2022) 053003},
  \href{http://arxiv.org/abs/2204.02886}{{\ttfamily arXiv:2204.02886
  [hep-ph]}}.

\bibitem{Geng:2007va}
C.~Q. Geng, S.~H. Ho, and J.~N. Ng, ``{Neutrino number asymmetry and
  cosmological birefringence},''
  \href{http://dx.doi.org/10.1088/1475-7516/2007/09/010}{{\em JCAP} {\bfseries
  09} (2007) 010}, \href{http://arxiv.org/abs/0706.0080}{{\ttfamily
  arXiv:0706.0080 [astro-ph]}}.

\bibitem{Ho:2010aq}
S.-H. Ho, W.~F. Kao, K.~Bamba, and C.~Q. Geng, ``{Cosmological birefringence
  due to CPT-even Chern-Simons-like term with Kalb-Ramond and scalar fields},''
  \href{http://dx.doi.org/10.1140/epjc/s10052-015-3426-5}{{\em Eur. Phys. J. C}
  {\bfseries 75} no.~5, (2015) 192},
  \href{http://arxiv.org/abs/1008.0486}{{\ttfamily arXiv:1008.0486 [hep-ph]}}.

\bibitem{Zhou:2023aqz}
R.-P. Zhou, D.~Huang, and C.-Q. Geng, ``{Cosmic birefringence from neutrino and
  dark matter asymmetries},''
  \href{http://dx.doi.org/10.1088/1475-7516/2023/07/053}{{\em JCAP} {\bfseries
  07} (2023) 053}, \href{http://arxiv.org/abs/2302.11140}{{\ttfamily
  arXiv:2302.11140 [astro-ph.CO]}}.

\bibitem{Horndeski:1976gi}
G.~W. Horndeski, ``{Conservation of Charge and the Einstein-Maxwell Field
  Equations},'' \href{http://dx.doi.org/10.1063/1.522837}{{\em J. Math. Phys.}
  {\bfseries 17} (1976) 1980--1987}.

\bibitem{Fleury:2014qfa}
P.~Fleury, J.~P. Beltran~Almeida, C.~Pitrou, and J.-P. Uzan, ``{On the
  stability and causality of scalar-vector theories},''
  \href{http://dx.doi.org/10.1088/1475-7516/2014/11/043}{{\em JCAP} {\bfseries
  11} (2014) 043}, \href{http://arxiv.org/abs/1406.6254}{{\ttfamily
  arXiv:1406.6254 [hep-th]}}.

\bibitem{Khodagholizadeh:2023aft}
J.~Khodagholizadeh, S.~Mahmoudi, R.~Mohammadi, and M.~Sadegh, ``{Cosmic
  birefringence as a probe of the nature of dark matter: Sterile neutrino and
  dipolar dark matter},''
  \href{http://dx.doi.org/10.1103/PhysRevD.108.023023}{{\em Phys. Rev. D}
  {\bfseries 108} no.~2, (2023) 023023},
  \href{http://arxiv.org/abs/2307.16286}{{\ttfamily arXiv:2307.16286
  [hep-ph]}}.

\bibitem{Greco:2022ufo}
A.~Greco, N.~Bartolo, and A.~Gruppuso, ``{Cosmic birefrigence: cross-spectra
  and cross-bispectra with CMB anisotropies},''
  \href{http://dx.doi.org/10.1088/1475-7516/2022/03/050}{{\em JCAP} {\bfseries
  03} no.~03, (2022) 050}, \href{http://arxiv.org/abs/2202.04584}{{\ttfamily
  arXiv:2202.04584 [astro-ph.CO]}}.

\bibitem{Greco:2022xwj}
A.~Greco, N.~Bartolo, and A.~Gruppuso, ``{Probing Axions through Tomography of
  Anisotropic Cosmic Birefringence},''
  \href{http://dx.doi.org/10.1088/1475-7516/2023/05/026}{{\em JCAP} {\bfseries
  05} (2023) 026}, \href{http://arxiv.org/abs/2211.06380}{{\ttfamily
  arXiv:2211.06380 [astro-ph.CO]}}.

\bibitem{Brezin:1971nd}
E.~Brezin and C.~Itzykson, ``{Polarization phenomena in vacuum nonlinear
  electrodynamics},'' \href{http://dx.doi.org/10.1103/PhysRevD.3.618}{{\em
  Phys. Rev. D} {\bfseries 3} (1971) 618--621}.

\bibitem{Grzadkowski:2010es}
B.~Grzadkowski, M.~Iskrzynski, M.~Misiak, and J.~Rosiek, ``{Dimension-Six Terms
  in the Standard Model Lagrangian},''
  \href{http://dx.doi.org/10.1007/JHEP10(2010)085}{{\em JHEP} {\bfseries 10}
  (2010) 085}, \href{http://arxiv.org/abs/1008.4884}{{\ttfamily arXiv:1008.4884
  [hep-ph]}}.

\bibitem{Lehman:2014jma}
L.~Lehman, ``{Extending the Standard Model Effective Field Theory with the
  Complete Set of Dimension-7 Operators},''
  \href{http://dx.doi.org/10.1103/PhysRevD.90.125023}{{\em Phys. Rev. D}
  {\bfseries 90} no.~12, (2014) 125023},
  \href{http://arxiv.org/abs/1410.4193}{{\ttfamily arXiv:1410.4193 [hep-ph]}}.

\bibitem{Liao:2020zyx}
Y.~Liao, X.-D. Ma, and Q.-Y. Wang, ``{Extending low energy effective field
  theory with a complete set of dimension-7 operators},''
  \href{http://dx.doi.org/10.1007/JHEP08(2020)162}{{\em JHEP} {\bfseries 08}
  (2020) 162}, \href{http://arxiv.org/abs/2005.08013}{{\ttfamily
  arXiv:2005.08013 [hep-ph]}}.

\bibitem{El-Menoufi:2015dra}
B.~K. El-Menoufi and G.~A. White, ``{The axial anomaly, dimensional
  regularization and Lorentz-violating QED},''
  \href{http://arxiv.org/abs/1505.01754}{{\ttfamily arXiv:1505.01754
  [hep-th]}}.

\bibitem{Heisenberg:1936nmg}
W.~Heisenberg and H.~Euler, ``{Consequences of Dirac's theory of positrons},''
  \href{http://dx.doi.org/10.1007/BF01343663}{{\em Z. Phys.} {\bfseries 98}
  no.~11-12, (1936) 714--732},
  \href{http://arxiv.org/abs/physics/0605038}{{\ttfamily
  arXiv:physics/0605038}}.

\bibitem{Schwinger:1951nm}
J.~S. Schwinger, ``{On gauge invariance and vacuum polarization},''
  \href{http://dx.doi.org/10.1103/PhysRev.82.664}{{\em Phys. Rev.} {\bfseries
  82} (1951) 664--679}.

\bibitem{Planck:2014ylh}
{\bfseries Planck} Collaboration, P.~A.~R. Ade {\em et~al.}, ``{Planck
  intermediate results - XXIV. Constraints on variations in fundamental
  constants},'' \href{http://dx.doi.org/10.1051/0004-6361/201424496}{{\em
  Astron. Astrophys.} {\bfseries 580} (2015) A22},
  \href{http://arxiv.org/abs/1406.7482}{{\ttfamily arXiv:1406.7482
  [astro-ph.CO]}}.

\bibitem{Ellis:2020unq}
J.~Ellis, M.~Madigan, K.~Mimasu, V.~Sanz, and T.~You, ``{Top, Higgs, Diboson
  and Electroweak Fit to the Standard Model Effective Field Theory},''
  \href{http://dx.doi.org/10.1007/JHEP04(2021)279}{{\em JHEP} {\bfseries 04}
  (2021) 279}, \href{http://arxiv.org/abs/2012.02779}{{\ttfamily
  arXiv:2012.02779 [hep-ph]}}.

\bibitem{Weinberg:1962zza}
S.~Weinberg, ``{Universal Neutrino Degeneracy},''
  \href{http://dx.doi.org/10.1103/PhysRev.128.1457}{{\em Phys. Rev.} {\bfseries
  128} (1962) 1457--1473}.

\bibitem{WMAP:2008lyn}
{\bfseries WMAP} Collaboration, E.~Komatsu {\em et~al.}, ``{Five-Year Wilkinson
  Microwave Anisotropy Probe (WMAP) Observations: Cosmological
  Interpretation},'' \href{http://dx.doi.org/10.1088/0067-0049/180/2/330}{{\em
  Astrophys. J. Suppl.} {\bfseries 180} (2009) 330--376},
  \href{http://arxiv.org/abs/0803.0547}{{\ttfamily arXiv:0803.0547
  [astro-ph]}}.

\bibitem{Altmannshofer:2018xyo}
W.~Altmannshofer, M.~Tammaro, and J.~Zupan, ``{Non-standard neutrino
  interactions and low energy experiments},''
  \href{http://dx.doi.org/10.1007/JHEP11(2021)113}{{\em JHEP} {\bfseries 09}
  (2019) 083}, \href{http://arxiv.org/abs/1812.02778}{{\ttfamily
  arXiv:1812.02778 [hep-ph]}}. [Erratum: JHEP 11, 113 (2021)].

\bibitem{Ellis:2018cos}
J.~Ellis and S.-F. Ge, ``{Constraining Gluonic Quartic Gauge Coupling Operators
  with gg\textrightarrow{}\ensuremath{\gamma}\ensuremath{\gamma}},''
  \href{http://dx.doi.org/10.1103/PhysRevLett.121.041801}{{\em Phys. Rev.
  Lett.} {\bfseries 121} no.~4, (2018) 041801},
  \href{http://arxiv.org/abs/1802.02416}{{\ttfamily arXiv:1802.02416
  [hep-ph]}}.

\bibitem{ATLAS:2017nga}
{\bfseries ATLAS} Collaboration, M.~Aaboud {\em et~al.}, ``{Search for dark
  matter at $\sqrt{s}=13$ TeV in final states containing an energetic photon
  and large missing transverse momentum with the ATLAS detector},''
  \href{http://dx.doi.org/10.1140/epjc/s10052-017-4965-8}{{\em Eur. Phys. J. C}
  {\bfseries 77} no.~6, (2017) 393},
  \href{http://arxiv.org/abs/1704.03848}{{\ttfamily arXiv:1704.03848
  [hep-ex]}}.

\bibitem{Mohammadi:2021xoh}
R.~Mohammadi, J.~Khodagholizadeh, M.~Sadegh, A.~Vahedi, and S.-s. Xue,
  ``{Cross-correlation power spectra and cosmic birefringence of the CMB via
  photon-neutrino interaction},''
  \href{http://dx.doi.org/10.1088/1475-7516/2023/06/044}{{\em JCAP} {\bfseries
  06} (2023) 044}, \href{http://arxiv.org/abs/2109.00152}{{\ttfamily
  arXiv:2109.00152 [hep-ph]}}.

\bibitem{Cremmer:1973mg}
E.~Cremmer and J.~Scherk, ``{Spontaneous dynamical breaking of gauge symmetry
  in dual models},'' \href{http://dx.doi.org/10.1016/0550-3213(74)90224-7}{{\em
  Nucl. Phys. B} {\bfseries 72} (1974) 117--124}.

\bibitem{Goldhaber:2008xy}
A.~S. Goldhaber and M.~M. Nieto, ``{Photon and Graviton Mass Limits},''
  \href{http://dx.doi.org/10.1103/RevModPhys.82.939}{{\em Rev. Mod. Phys.}
  {\bfseries 82} (2010) 939--979},
  \href{http://arxiv.org/abs/0809.1003}{{\ttfamily arXiv:0809.1003 [hep-ph]}}.

\bibitem{ParticleDataGroup:2022pth}
{\bfseries Particle Data Group} Collaboration, R.~L. Workman {\em et~al.},
  ``{Review of Particle Physics},''
  \href{http://dx.doi.org/10.1093/ptep/ptac097}{{\em PTEP} {\bfseries 2022}
  (2022) 083C01}.

\bibitem{Nieves:1988qz}
J.~F. Nieves and P.~B. Pal, ``{$P$ and {CP} Odd Terms in the Photon Self-energy
  Within a Medium},'' \href{http://dx.doi.org/10.1103/PhysRevD.39.652}{{\em
  Phys. Rev. D} {\bfseries 39} (1989) 652}. [Erratum: Phys.Rev.D 40, 2148
  (1989)].

\bibitem{Giunti:2014ixa}
C.~Giunti and A.~Studenikin, ``{Neutrino electromagnetic interactions: a window
  to new physics},'' \href{http://dx.doi.org/10.1103/RevModPhys.87.531}{{\em
  Rev. Mod. Phys.} {\bfseries 87} (2015) 531},
  \href{http://arxiv.org/abs/1403.6344}{{\ttfamily arXiv:1403.6344 [hep-ph]}}.

\bibitem{Jenkins:2017jig}
E.~E. Jenkins, A.~V. Manohar, and P.~Stoffer, ``{Low-Energy Effective Field
  Theory below the Electroweak Scale: Operators and Matching},''
  \href{http://dx.doi.org/10.1007/JHEP03(2018)016}{{\em JHEP} {\bfseries 03}
  (2018) 016}, \href{http://arxiv.org/abs/1709.04486}{{\ttfamily
  arXiv:1709.04486 [hep-ph]}}.

\end{thebibliography}
\end{document}